\documentstyle[11pt]{article}
\if@twoside\oddsidemargin25mm
\else\oddsidemargin25mm\evensidemargin25mm\marginparwidth25mm\fi
\def\thefootnote{\dag}
\def\bl{\Big\{}
\def\br{\Big\}}
\def\bpl{\Big(}
\def\bpr{\Big)}
\def\ve{\varepsilon}
\def\t{\theta}
\def\vphi{\varphi}
\def\w{\omega}
\def\O{\Omega}
\def\F{{\cal F}}

\def\der{\partial}
\def\bq{\begin{equation}}
\def\eq{\end{equation}}
\def\brr{\begin{eqnarray}}
\def\err{\end{eqnarray}}
\def\ba{\left(\begin{array}}
\def\ea{\end{array}\right)}
\def\dslash{\hbox{\ooalign{$\displaystyle\partial$\cr$/$}}}
\def\Dslash{\hbox{\ooalign{$\displaystyle D$\cr$\hspace{.03in}/$}}}

\def\Eslash{\hbox{\ooalign{$\displaystyle E$\cr$\hspace{.03in}/$}}}

\def\hatVslash{\hbox{\ooalign{$\displaystyle {\hat V}$\cr$
\hspace{.02in}/$}}}

\def\derbar{\stackrel{\leftrightarrow}{\partial}}
\newcommand{\dr}{\raise.3ex\hbox{$\stackrel{\leftarrow}{\partial }$}{}}
\newcommand{\dl}{\raise.3ex\hbox{$\stackrel{\rightarrow}{\partial}$}{}}
\newcommand{\eqn}[1]{(\ref{#1})}
\newcommand{\ft}[2]{{\textstyle\frac{#1}{#2}}}

\begin{document}
\renewcommand{\a}{\alpha}
\renewcommand{\b}{\beta}
\renewcommand{\c}{\gamma}
\renewcommand{\d}{\delta}
\newcommand{\pa}{\partial}
\newcommand{\g}{\gamma}
\newcommand{\G}{\Gamma}
\newcommand{\A}{\Alpha}
\newcommand{\B}{\Beta}
\newcommand{\D}{\Delta}
\newcommand{\e}{\epsilon}
\newcommand{\E}{\Epsilon}
\newcommand{\z}{\zeta}
\newcommand{\Z}{\Zeta}
\newcommand{\K}{\Kappa}
\renewcommand{\l}{\lambda}
\renewcommand{\L}{\Lambda}
\newcommand{\La}{\Lambda}
\newcommand{\m}{\mu}
\newcommand{\M}{\Mu}
\newcommand{\n}{\nu}
\newcommand{\N}{\Nu}
\newcommand{\x}{\chi}
\newcommand{\X}{\Chi}
\newcommand{\p}{\pi}
\newcommand{\R}{\Rho}
\newcommand{\s}{\sigma}
\renewcommand{\S}{\Sigma}
\newcommand{\T}{\Tau}
\newcommand{\y}{\upsilon}
\newcommand{\Y}{\upsilon}
\renewcommand{\o}{\omega}
\newcommand{\q}{\theta}
\newcommand{\h}{\eta}
\begin{titlepage}
\begin{flushright} KUL-TF-97/24\\ THU-97/26\\ HUB-EP-97/73 \\
ULB-TH-97/18\\ ITP-SB-97-63 \\ DFTT-62/97 \\[3mm] 
{\tt hep-th/9710212}
\end{flushright}
\vspace{16mm}
\begin{center}
{\LARGE\bf N=2 Supergravity Lagrangians with \\[.1in]
Vector-Tensor Multiplets}                                 
\vfill
{\large P. Claus$^1$,
        B. de Wit$^2$, M. Faux$^{1,3}$\\[1.5mm]
        B. Kleijn$^2$, R. Siebelink$^{1,4,5}$ and P. Termonia$^6$}\\
\vspace{7mm}
{\small
$^1$ Instituut voor Theoretische Fysica - Katholieke Universiteit Leuven\\
     Celestijnenlaan 200D, B-3001 Leuven, Belgium\\[6pt]
$^2$ Institute for Theoretical Physics - Utrecht University\\
     Princetonplein 5, 3508 TA Utrecht, The Netherlands\\[6pt]
$^3$ Humboldt-Universit\"at zu Berlin, Institut f\"ur Physik\\
     D-10115 Berlin, Germany \\[6pt]
$^4$ Service de Physique Th\'{e}orique, Universit\'{e} Libre de
     Bruxelles, \\ 
     Campus Plaine, CP 225, Bd du Triomphe, B-1050 Bruxelles, Belgium\\[6pt]
$^5$ Institute for Theoretical Physics, State University of New 
     York at Stony Brook\\ 
     Stony Brook, NY 11790-3840, USA\\[6pt]
$^6$ Dipartimento di Fisica Teorica, Universit\'a di Torino\\ 
     via P. Giuria 1, I-10125 Torino\\
     Istituto Nazionale di Fisica Nucleare (INFN) -- Sezione di 
     Torino, Italy  }
\end{center}
\vfill
\begin{center} {\bf Abstract} \\[3mm]
{\small
We discuss the coupling of vector-tensor 
multiplets to $N=2$ supergravity. }
\end{center}
\vfill
\flushleft{October 1997}
\end{titlepage}
\renewcommand\thefootnote{\arabic{footnote}}
\section{Introduction and summary}
Theories with extended supersymmetry have provided a rich testing 
ground  for the study of many phenomena in field 
theory and string theory \cite{SW}.  
This paper deals with $N=2$ supergravity in four-dimensional 
spacetimes, whose coupling to a number of matter multiplets has 
been worked out in considerable   
detail. The most well-known matter multiplets are the vector 
multiplet and the hypermultiplet. Off-shell a vector multiplet 
comprises $8+8$  
bosonic and fermionic degrees of freedom \cite{grimm}. The off-shell 
representation of hypermultiplets \cite{fayet} is more subtle. 
To remain off-shell one can either 
choose a formulation with (off-shell) central charges based on 
$8+8$ degrees of freedom and ensuing constraints, or one 
must accept a description based on an infinite number of degrees
of freedom. For the latter description,  
harmonic superspace provides a natural setting \cite{harmonic}. 
The tensor multiplet \cite{DWVH}, which is dual to a massless 
hypermultiplet, also  
comprises $8+8$ off-shell degrees of freedom. All three 
multiplets describe $4+4$ physical degrees of freedom. 

There are two other matter multiplets describing the same 
number of physical states. First there is the 
vector-tensor multiplet \cite{sohnius,DWKLL}, which is dual to a 
vector multiplet. 
Secondly, there exists a double-tensor multiplet, which 
contains two tensor gauge fields and which is dual to a 
hypermultiplet. The off-shell representation of both these 
multiplets requires the presence of off-shell central charges and 
their superconformal  
formulation requires the presence of background fields. 
All multiplets appear naturally in the context of 
four-dimensional $N=2$ supersymmetric 
compactifications of the three ten-dimensional supergravity 
theories, and may therefore play a role in the effective 
low-energy actions associated with appropriate string 
compactifications. For instance, the vector-tensor 
multiplet can be associated with the supermultiplet of vertex 
operators of four-dimensional heterotic $N=2$ supersymmetric 
string vacua, which contains the operators of 
the axion-dilaton complex together with an extra vector gauge 
field. Similarly, the tensor and the double-tensor multiplet 
would appear in the context of type-IIA and type-IIB string  
compactifications.  

At the level of the four-dimensional effective actions, the 
latter multiplets are  
usually converted into vector multiplets and hypermultiplets, 
which, at least in string-perturbation theory, yields an 
equivalent description. We should stress that this conversion 
rests on a purely on-shell equivalence. The question whether certain 
off-shell configurations are prefered by string theory has a 
long history (for a recent discussion, see \cite{siegel}). 
At any rate, not every system of vector multiplets or 
hypermultiplets can be  converted back into vector-tensor, tensor 
or double-tensor multiplets, so there are certain restrictions 
(see, e.g. \cite{siebel}). Furthermore, recent  experience in  
dual systems, for instance in the context of three spacetime 
dimensions \cite{3dmirror},  has taught us that the answer to 
these questions involves nonperturbative issues. While in 
\cite{DWKLL} the vector-tensor multiplet was introduced, motivated 
by heterotic string perturbation theory, it meanwhile turned out 
that vector-tensor multiplets have a different role to play and 
emerge in heterotic compactifications at the nonperturbative 
level. This phenomenon was initially described in the context 
of six-dimensional heterotic string compactifications, where it 
turned out that certain singularities in the effective action 
were associated with noncritical strings becoming tensionless 
\cite{SeiWit}. In six-dimensions this is related to the presence 
of tensor multiplets. In four dimensions, vector-tensor 
multiplets play a similar role \cite{LSTY}. Couplings of 
the vector-tensor  
multiplet appear in two varieties. One type of coupling could be 
of six-dimensional origin \cite{FerMinSag}. The  
origin of the second coupling is less clear. For  recent 
discussions on the couplings of six-dimensional tensor multiplets, 
we refer to \cite{BSS}. 

In two previous publications \cite{vt1,vt2} we have developed 
the coupling of  
the vector-tensor multiplet in the context of the superconformal 
multiplet calculus. So far we restricted ourselves to rigid 
supersymmetry and we constructed all possible 
couplings to a general (background) configuration of vector 
multiplets that are 
invariant under rigid scale and chiral U(1) transformations. The 
requirement of scale and chiral invariance forces the scalar 
fields of the vector multiplet to act  
as compensators. In the context of rigid supersymmetry this 
feature does not represent a restriction: at the end one can 
always freeze the vector multiplets to constants, thereby causing 
a breaking of scale invariance. In the case of local 
supersymmetry it is more subtle to freeze the vector 
multiplets. The reason for insisting on rigid scale and chiral 
invariance, is that, in this form,  one can rather 
straightforwardly 
incorporate the coupling to supergravity by employing the 
superconformal multiplet calculus \cite{DWVHVP,DWLVP}. In this 
paper, we report on the results of the extension to local 
supersymmetry. We give a 
comprehensive treatment of matter couplings to $N=2$  
supergravity. For the vector-tensor multiplet we make use of a 
formulation based on a finite  
number of off-shell degrees of freedom, which employs off-shell 
central charges. These degrees of freedom are described by one  
vector, one two-rank tensor gauge field, two real scalar fields, 
one of them auxiliary, and a doublet of Majorana spinors. In this 
formulation the gauge field associated  
with the central charge is known to exhibit rather peculiar 
couplings \cite{zachos,DWVHVP}. Recently another specific example of
such a coupling was studied in \cite{BD}. 

In \cite{vt2} we established the existence of two different vector-tensor 
multiplets. Their difference is encoded in the Chern-Simons 
couplings between the vector and tensor gauge fields, 
whose form is constrained by supersymmetry.
One version, first discussed in \cite{vt1}, is characterized by 
the fact that the vector and tensor 
gauge field of the vector-tensor multiplet exhibit a direct Chern-Simons 
coupling. This leads to unavoidable nonlinearities (in terms of 
the vector-tensor multiplet fields) of the action 
and transformation rules. This theory is formulated with at least 
one abelian vector multiplet, which provides the gauge field for 
the central-charge transformations. When freezing this vector 
multiplet to a constant we obtain a vector-tensor multiplet with 
a self interaction. It takes the form (after 
a suitable rescaling of the fields)
\brr\label{vt-selfinteraction}
{\cal L}&\propto&  \ft12 \phi (\der_\m\phi)^2  + \ft14  
\phi  (\der_\m V_\n-\der_\n V_\m)^2 +\ft34 \phi^{-1} 
\Big(\der_{[\m}B_{\n\rho]}- V_{[\m}\der_\n V_{\rho]}\Big)^2 \nonumber \\
&& +\ft12 \phi\, \bar \l^i {\stackrel{\leftrightarrow}{\dslash}} \l_i 
-2\phi\, (\phi^{({\rm z})})^2- \ft14 i\Big(\ve^{ij} \bar 
\l_i\s^{\m\n} \l_j - {\rm h.c.}\Big) \, 
(\der_\m V_\n - \der_\n V_\m) \nonumber\\
&& -\ft1{24} \phi^{-1} \bar \l_i \g^\m \g^\n\g^\rho  
\l^i  \Big(\der_{[\m} B_{\n\rho]} - V_{[\m} \der_\n V_{\rho]} \Big)
 + \ft3{16} \phi^{-1} \Big(\bar\l^i\g_\m\l_i\Big)^2 \nonumber\\
&&  + \ft1{32} \phi^{-1}\Big((\ve^{ij} \bar\l_i\s_{\m\n}\l_j)^2 + {\rm 
h.c.}  \Big) \,,
\err
where we have included an auxiliary field, $\phi^{({\rm z} )}$. We 
will comment on its role in due course. The second version of the 
vector-tensor multiplet, which requires 
more than one abelian vector multiplet, avoids the direct Chern-Simons 
coupling between the vector and tensor field of the vector-tensor 
multiplet, but there are nonvanishing Chern-Simons couplings with 
the additional vector multiplets. In this case the action 
remains quadratic in terms of the vector-tensor multiplet fields. 

Recently a number of papers appeared dealing with the superspace 
formulation of vector-tensor multiplets \cite{GHH,HOW,DK,BHO,DK2}. 
Most of this work concerns the linear version of the 
vector-tensor multiplet with its corresponding Chern-Simons 
couplings, which 
can be obtained by dimensional reduction from six dimensions \cite{BSS}. 
Unfortunately, even in the framework of harmonic superspace, 
it turns out that it is not possible to avoid an explicit 
central charge with corresponding constraints \cite{DK}. On the other hand, 
the complexity of our results clearly demonstrates the 
need for a suitable superspace formulation. For rigid
supersymmetry, the self-interaction  
\eqn{vt-selfinteraction} has been derived recently in harmonic 
superspace \cite{DK2,IS}.  

This paper is organized as follows. In section~2 we present a
survey of the superconformal multiplet calculus and establish 
our notation. In section~3 we introduce the vector-tensor 
multiplet and discuss its superconformal transformation rules. 
Section~4 contains the derivation of the locally supersymmetric 
actions for vector-tensor multiplets. In section~5 we discuss 
their dual version in terms of vector multiplets. A number of useful 
formulae has been collected in an appendix.

\setcounter{equation}{0}
\section{Superconformal Multiplet Calculus}
Off-shell formulations of supergravity theories
can be described in a form that is gauge equivalent to a 
superconformal theory. In four spacetime dimensions this enables 
a relatively concise organization 
of the field content.  It also allows the systematic construction of
supersymmetric Lagrangians via techniques
known collectively as multiplet calculus.
In this section we review these concepts for the case of $N=2$
theories to make this paper self-contained and to establish our 
notation. Most of the material presented here is known (see, 
e.g., \cite{DWVHVP,DWLVP}).  

One is interested, firstly, in identifying irreducible representations
of the superconformal algebra. The relevant algebra contains 
general-coordinate, local Lorentz ($M$), dilatation ($D$),
special conformal ($K$), chiral SU(2) and U(1), supersymmetry 
($Q$) and special supersymmetry ($S$) transformations. The most 
conspicuous aspects of this algebra are that it involves field-dependent 
structure 
`constants' and that only a subset of the gauge fields are realized 
as independent fields. To be specific, the gauge fields associated with 
general-coordinate transformations ($e_\m^a$), dilatations ($b_\m$), 
chiral symmetry (${\cal V}^{\,i}_{\m\, j}, A_\m$) 
and $Q$-supersymmetry ($\psi_\m^i$), are realized by independent fields. The 
remaining gauge fields of Lorentz ($\o^{ab}_\m$), special 
conformal ($f^a_\m$) and $S$-supersymmetry transformations
($\phi_\m^i$) are dependent fields.  
Their form is determined by a set of covariant constraints.
The identification of the
appropriate constraints and the precise commutator relations is
nontrivial \cite{DWVHVP}. Of primary interest is
the Weyl multiplet, which is the representation consisting of 
$24+24$ off-shell degrees of freedom, corresponding to the
independent gauge fields associated with the superconformal 
algebra and three auxiliary fields: a Majorana spinor doublet 
$\chi^i$, a scalar $D$ and a selfdual Lorentz tensor $T_{abij}$ (where $i,
j,\ldots$ are chiral SU(2) spinor indices)\footnote{%
  $T_{abij}$ is antisymmetric in both Lorentz indices $a,b$ and 
  chiral SU(2) indices $i,j$. It is a selfdual Lorentz tensor and 
  therefore complex. Its complex conjugate is the anti-selfdual 
  field $T^{ij}_{ab}$. Obviously the tensor field transforms as a 
  singlet under SU(2), but it transforms nontrivially under 
  chiral U(1). Our conventions are such that SU(2) indices are 
raised and lowered by complex conjugation. The SU(2) gauge field 
${\cal V}_\m^{\;i}{}_j$ is antihermitean and traceless, i.e., 
${\cal V}_\m^{\;i}{}_j+{\cal V}_{\m j}{}^{i}= {\cal 
V}_\m^{\;i}{}_i=0$. We refer to the 
appendix for further details. }. %

When additional supermultiplets, such as  vector multiplets or
vector-tensor multiplets,  are added to the superconformal theory,
additional gauge symmetries may arise, which must be included into
the algebra. We denote these extra symmetry transformations by
$\d_{{\rm gauge}}$, which may incorporate 
central-charge transformations. The Weyl multiplet itself is invariant under
$\d_{{\rm gauge}}$. The most important of the commutator relations
which specify the algebra, is the one between a pair of $Q$-supersymmetry
transformations, given by
\bq [\d_Q(\e_1),\d_Q(\e_2)]=
    \d^{(cov)}(\xi)
    +\d_M(\ve)+\d_K(\La_K)+\d_S(\h)+\d_{{\rm gauge}} \,,
\label{qqcomb}\eq
where $\d^{(cov)}, \d_M, \d_K$ and $\d_S$ denote a covariant 
general-coordinate transformation, a Lorentz transformation,
a special conformal transformation, and an $S$-supersymmetry
transformation. The associated parameters are given by the 
following expressions, 
\brr \xi^\mu &=& 2\,\bar{\e}_2^i\g^\mu\e_{1i}+{\rm h.c.}\,, \nonumber\\
     \ve^{ab} &=& \bar{\e}^i_1\e^j_2\,T^{ab}_{ij}+{\rm h.c.}\,, \nonumber\\
     \La_K^a &=& \bar{\e}^i_1\e^j_2\, D_bT^{ba}_{ij}
     -\ft32\bar{\e}_2^i\g^a\e_{1i}\,D+{\rm h.c.}\,, \nonumber\\
     \eta^i &=& 3\,\bar{\e}^i_{[1}\e^j_{2]}\,\chi_j  \,,
\label{qqparamsb}\err
where $D_b$ denotes the derivative that is covariant with 
respect to all the superconformal symmetries. 
In the sequel we
will occasionally need the commutator of an $S$- and a 
$Q$-supersymmetry variation\footnote{%
  To clarify our notation, for instance, $\bar{\h}^i\e_j-({\rm h.c.}\,
  ;\,{\rm traceless}) =\bar{\h}^i\e_j-\bar{\h}_j\e^i
  -\frac{1}{2}\d^i{}_{\!j}(\bar{\h}^k\e_k-\bar{\h}_k\e^k)$.} 
 :%
\brr 
[\d_S(\h), \d_Q(\e)] &=& \d_M \bpl 2 \bar{\h}^i\s^{ab}\e_i + {\rm h.c.} \bpr
+ \d_D \bpl \bar{\h}_i \e^i + {\rm h.c.} \bpr
+ \d_{\rm U(1)} \bpl  i\bar{\h}_i \e^i + {\rm h.c.} \bpr \nonumber\\
&& + \d_{\rm SU(2)} \bpl -2  \bar{\h}^i \e_j -({\rm h.c.}\,;\,
{\rm traceless}) 
\bpr \,.
\label{SQcomm}
\err
Given the $S$-supersymmetry variations one may compute the 
special conformal boosts from the commutator
\bq
[\d_S(\h_1),\d_S(\h_2)] = \d_K (\La^a_K) \,,\quad \mbox{ with } 
\L^a_K = \bar \eta_{2i}\g^a\eta^i_1 + {\rm h.c.}\,.
\eq

Poincar\'e supergravity theories are obtained by coupling the Weyl multiplet
to additional superconformal multiplets containing Yang-Mills and
matter fields. The resulting superconformal theory then becomes 
gauge equivalent to a theory of Poincar\'e supergravity. This is 
conveniently exploited by imposing gauge conditions on certain 
components of the extra superconformal multiplets. Subsequently 
one can eliminate the auxiliary superconformal fields.
The additional multiplets are necessary to provide
compensating fields and to overcome a deficit in degrees of freedom
between the Weyl multiplet and the minimal field representation of
Poincar\'e supergravity.  For instance, the graviphoton, represented by
an abelian vector field in the Poincar\'e supergravity multiplet,
is provided by an $N=2$ superconformal vector multiplet.

In the following subsections we briefly describe the Weyl multiplet, 
vector multiplets, hypermultiplets and linear multiplets.
\vspace{.1in}

\subsection{The Weyl multiplet}
We already specified the fields belonging to the Weyl multiplet. 
The Weyl and chiral weights and the fermion chiralities of the 
Weyl-multiplet fields, the composite connections, and also those of the
supersymmetry transformation parameters, are shown in table 2.1. 
The Weyl and chiral weights, $w$ and $c$, govern the 
transformation of a generic field under dilatations and U(1) 
transformations according to 
\bq
\phi(x) \longrightarrow \exp[ w \,\L_{\rm D} (x) + i c\,\L_{\rm 
U(1)}(x) ]\,\phi(x)\,.
\eq
Here we summarize the transformation rules for the independent 
fields under $Q$- and $S$-super\-symmetry and under 
$K$-transformations, 
\brr \d e_\mu{}^a &=&
     \bar{\e}^i\g^a\psi_{\mu i}+{\rm h.c.}\,, \nonumber\\
      \d\psi_\mu^i &=& 2{\cal D}_\mu\e^i
      -\ft14 \s\cdot T^{ij}\g_\mu\e_j
      -\g_\mu\eta^i \,,\nonumber\\
      \d b_\mu &=&
      \ft12\bar{\e}^i\phi_{\mu i}
      -\ft34\bar{\e}^i\g_\mu\chi_i
      -\ft12\bar{\eta}^i\psi_{\mu i}+{\rm h.c.} +\L_K^a\,e_\m^a  
      \,,\nonumber\\   
      \d A_\mu &=& \ft{1}{2}i \bar{\e}^i\phi_{\mu i}
      +\ft{3}{4}i\bar{\e}^i\g_\mu\chi_i
      +\ft{1}{2}i \bar{\eta}^i\psi_{\mu i}+{\rm h.c.}\,, \nonumber\\
      \d {\cal V}_{\mu\,j}^i &=&
      2\bar{\e}_j\phi_\mu^i-3\bar{\e}_j\g_\mu\chi^i+2\bar{\eta}_j\psi_\mu^i
     -({\rm h.c.} \, ; \, {\rm traceless})\,,
      \nonumber\\
      \d T_{ab}^{ ij} &=&
      8 \bar{\e}^{[ i} \hat{R}_{ab}(Q)^{j]}\,, \nonumber\\
      \d\chi^i &=& -\ft16\s\cdot\Dslash T^{ij}\e_j
      +\ft{1}{3}\hat{R}({\rm SU(2)})^i{}_{j}\cdot\s\e^j
      -\ft{2}{3}i \hat{R}({\rm U(1)})\cdot\s\e^i \nonumber\\
      && +D\,\e^i
      +\ft16\s\cdot T^{ij}\eta_j \,,\nonumber\\
      \d D &=& \bar\e^i\Dslash\chi_i+{\rm h.c.}\,,
 \label{transfo4}\err
where ${\cal D}_\mu$ are derivatives covariant with respect to
Lorentz, dilatational, U(1) and SU(2) transformations,
and $D_\mu$ are derivatives covariant with respect to {\it all}
superconformal transformations. Both ${\cal D}_\m$ 
and $D_\m$ are covariant with respect to the additional gauge transformations 
associated with possible gauge fields of the matter multiplets. 
The quantities $\hat{R}_{ab}(Q), 
\hat{R}_{ab}({\rm U(1)})$ and $\hat{R}_{ab}({\rm SU(2)})^i_{\,j}$ 
are supercovariant curvatures 
related to $Q$-supersymmetry, U(1) and SU(2) transformations.
Their precise definitions are given in the appendix.
The gauge fields for Lorentz, special conformal, and $S$-supersymmetry
transformations are denoted $\w_\mu^{ab}, \phi_\mu^i$,
and $f_\mu^a$, respectively.  These are composite objects, which 
depend in a complicated way on the independent fields (see the appendix).
Under supersymmetry and special conformal boosts they transform as follows,
\brr \d\w_\mu^{ab} &=& -\bar{\e}^i\s^{ab}\phi_{\mu i} 
     -\ft12\bar{\e}^iT^{ab}_{ij}\psi_\mu^j
     +\ft32\bar{\e}^i\g_\mu\s^{ab}\chi_i \nonumber\\
     && +\bar{\e}^i\g_\mu\hat{R}^{ab}(Q)_i 
    -\bar{\eta}^i\s^{ab}\psi_{\mu i} + {\rm h.c.}  + 
    2\L_K^{[a}\,e_\m^{b]} \,, \nonumber\\
     \d\phi_\mu^i &=& -2f_\mu^a\g_a\e^i
     -\ft14\Dslash T^{ij}\cdot\s\g_\mu\e_j
     +\ft32\big[(\bar{\chi}_j\g^a\e^j)\g_a\psi_\mu^i
     -(\bar{\chi}_j\g^a\psi_\mu^j)\g_a\e^i\big] \nonumber\\
     & & +\ft{1}{2}\hat{R}({\rm SU(2)})^i{}_{j} \cdot\s\g_\mu\e^j
     +i\hat{R}({\rm U(1)})\cdot\s\g_\mu\e^i +2{\cal D}_\mu\eta^i 
     +\L^a_K \g_a\psi_\m^i \,,\nonumber\\
     \d f_\mu^a &=& -\ft12\bar{\e}^i\psi_\mu^j\, D_b T^{ba}_{ij}
     -\ft34e_\mu{}^a\bar{\e}^i\Dslash\chi_i
     -\ft34\bar{\e}^i\g^a\psi_{\mu i}\,D 
       \nonumber\\
     & & +\bar{\e}^i\g_\mu D_b\hat{R}^{ba}(Q)_i 
     +\ft12\bar{\eta}^i\g^a\phi_{\mu i}+ {\rm h.c.} +{\cal 
     D}_\mu\L^a_K \,.
\err
\begin{figure}
\begin{center}
\begin{tabular}{|c||cccccccc|ccc||cc|}
\hline
&
   \multicolumn{11}{c||}{Weyl multiplet} &
   \multicolumn{2}{c|}{parameters} \\
\hline
\hline
field          &
   $e_\mu{}^a$   &
   $\psi_\mu^i$  &
   $b_\mu$       &
   $A_\mu$       &
   ${{\cal V}_\mu}^i{}_j$ &
   $T_{ab}^{ij}$    &
   $\chi^i$      &
   $D$           &
   $\w_\mu^{ab}$ &
   $f_\mu{}^a$   &
   $\phi_\mu^i$  &
   $\e^i$        &
   $\eta^i$ \\[.5mm]
\hline
\hline
$w$         & $-1$     & $-\ft12$  & $0$      & $0$      & $0$    
  & $1$      &  $\ft32$  & $2$      & $0$       & $1$      & 
$\ft12$  & $-\ft12$ & $\ft12$ \\[.5mm] 
 \hline
$c$         & $0$      & $-\ft12$  & $0$      & $0$      & $0$    
  & $-1$     &  $-\ft12$ & $0$      & $0$       & $0$      & 
$-\ft12$ & $-\ft12$ & $-\ft12$ \\[.5mm] 
\hline
$\gamma_5$&          & $+$       &          &          &          
&          & $+$       &          &           &          &    $-$ 
  &  $+$     &  $-$     \\[.5mm] 
\hline
\end{tabular}\\[.13in]
\parbox{5.7in}{Table 2.1: Weyl and chiral weights ($w$ and $c$, respectively)
             and fermion chirality ($\gamma_5$)
             of the Weyl multiplet component fields
             and of the supersymmetry transformation parameters.}
\end{center}
\end{figure}

\subsection{The vector multiplet}
The $N=2$ vector multiplet transforms in the adjoint representation
of a given gauge group. For each value of the group index $I$,
there are $8+8$ component degrees of freedom off-shell,
including a complex scalar $X^{I}$, a doublet of chiral fermions $\O_i^{\,I}$,
a vector gauge field $W_\mu^{\,I}$,
and a real $SU(2)$ triplet of scalars\footnote{%
   The real triplet $Y_{ij}^{\,I}$ satisfies  $Y_{ij}^{\,
   I}=Y_{ji}^{\,I}$ and $Y_{ij}^{\, I}=\ve_{ik}\ve_{jl}Y^{kl\,I}$.} %
$Y_{ij}^{\,I}$. The Weyl and chiral weights and the fermion chirality of the
vector-multiplet component fields are listed in table 2.2.
Under $Q$- and $S$-supersymmetry these transform as follows,
\brr \d X^{I} &=& \bar{\e}^i\O_i^{\,I} \,,\nonumber\\
     \d\O_i^{\,I} &=& 2\Dslash X^{I}\e_i
     +\ve_{ij}\s\cdot \F^{I-}\e^j
     +Y_{ij}^{\,I}\e^j
     -2gf_{JK}{}^IX^{J}\bar{X}^{K}\ve_{ij}\e^j
     +2X^{I}\eta_i\,,
     \nonumber\\
     \d W_\mu^I &=& \ve^{ij}\bar{\e}_i\g_\mu\O_j^{\,I}
     +2\ve_{ij}\bar{\e}^i\bar{X}^{I}\psi_\mu^j+ {\rm h.c.}\,, \nonumber\\
     \d Y_{ij}^{\,I} &=& 2\bar{\e}_{(i}\Dslash\O_{j)}^I
     +2\ve_{ik}\ve_{jl}\bar{\e}^{(k}\Dslash\O^{l)\,I}
     -4gf_{JK}{}^I\, \ve_{k(i}\Big(\bar{\e}_{j)}X^J\O^{k\,K}
     -\bar{\e}^k\bar{X}^J\O_{j)}^{\,K}\Big) \,, 
\label{vrules}\err
where $f_{JK}{}^I$ are the structure constants of the group,
$[t_I,t_J]=f_{IJ}{}^K\,t_K$, and
$g$ is a coupling constant. The field strengths $\F_{\mu\nu}^I$
are defined by
\bq \F_{\mu\nu}^I=
    2\der_{[\mu}W_{\nu]}^I
    -g f_{JK}{}^I\, W_\mu^J W_\nu^K
    -\Big(\ve_{ij}\bar{\psi}_{[\mu}^i\g_{\nu]}\O^{j\,I}
    +\ve_{ij}\bar{X}^{I}\bar{\psi}_\mu^i\psi_\nu^j
    +\ft14\ve_{ij}\bar{X}^{I}T^{ij}_{\mu\nu}+{\rm h.c.}\Big) \,.
\label{calF}
\eq
They satisfy the Bianchi identity
\bq
D^b\Big(\F^{+I}_{ab} -  \F^{-I}_{ab} +\ft14 X^I T_{ab\,ij} 
\varepsilon^{ij} -\ft14 \bar X^I T^{ij}_{ab} \varepsilon_{ij} \Big) = 
\ft34 \Big(\bar \chi^i\g_a \O^{Ij}\varepsilon_{ij} -\bar \chi_i\g_a 
\O_j^I\varepsilon^{ij}  \Big)\,. 
\eq
Under supersymmetry they transform as follows,
\bq \d\F^{I}_{ab}=
    -2 \ve^{ij}\bar{\e}_i\g_{[a}D_{b]}\O_j^{\,I}
    -2\ve^{ij}\bar{\eta}_i\s_{ab}\O_j^{\,I} + {\rm h.c.}\,.
 \eq
The transformation rules (\ref{vrules}) satisfy the commutator
relation (\ref{qqcomb}), including a field-dependent gauge transformation
on the right-hand side, which acts with the following parameter
\bq
\t^I=4\ve^{ij}\bar{\e}_{2i}\e_{1j}\,X^I+
 {\rm h.c.} \,.
\label{vgauge}
\eq

The covariant quantities of the vector multiplet constitute a so-called 
reduced chiral multiplet. A general chiral multiplet
contains $16+16$ off-shell degrees of freedom
and an arbitrary Weyl weight factor $w$ (corresponding to the
Weyl weight of its lowest component). The covariant quantities of 
the vector multiplet may be obtained
from a chiral multiplet with $w=1$ by the application of a set of
reducibility conditions, one of which is the Bianchi identity.
\vspace{.1in}

\subsection{The hypermultiplet} 
A finite field configuration describing off-shell hypermultiplets
must have a nontrivial central charge. This charge acts on a
basic unit underlying $r$  hypermultiplets, which consists of
$r$ quaternions $A_i^{\;\a}$ and $2r$ chiral fermions $\zeta^\a$.
The Weyl and chiral weights and fermion chirality of these fields
are listed in table 2.2.\footnote{%
   Our notation is such that the $2\times 2r$ matrix
   $A_i^{\;\a}$, with complex conjugate $A^i_{\;\a}$,
   satisfies the constraint $A^i_{\;\a} =
   \ve^{ij}\rho_{\a\b}A_j^{\;\b}$
   where, under certain conditions \cite{DWLVP},  $\rho_{\a\b}$ can be 
   brought in block-diagonal form, 
   $\rho ={\rm diag}(i\s_2, i\s_2,...)$. Solving this constraint
   reduces $A^i_{\;\a}$ to a sequence of $r$ quaternions $(q_1,...q_r)$,
   where each quaternion is represented by the $2\times 2$ matrix
   $q_a=q_a^{(0)}+iq_a^{(1)}\s_1+iq_a^{(2)}\s_2+iq_a^{(3)}\s_3$.}
The index $\a$ runs from $1$ to $2r$. As the basic unit contains
twice as many 
fermionic as bosonic components, it is necessary to assume the presence of 
an infinite number of them. These multiple ``copies'', which will
be distinguished by appending successive ``z'' indices to the
fields, will be
organized in a linear chain, such that the central charge maps each one of
them into the next one. For instance, on $A_i{}^\a$ it acts as
$\d A_i = z A_i^{\;\a({\rm z})}$, where $z$ is the transformation parameter. 
Successive applications
of the central charge thus generate an infinite sequence,
\bq
A_i{}^\a\longrightarrow A_i{}^{\a({\rm z})}\longrightarrow 
A_i{}^{\a({\rm zz})}\longrightarrow
{\rm etcetera}\,,
\label{hhier}
\eq
and similarly on the fermionic fields\footnote{%
   A hierarchy such as (\ref{hhier}) arises naturally when starting
   from a five-dimensional supersymmetric theory with one compactified
   coordinate, but this interpretation is not essential.}.
The supersymmetry transformation rules for the basic fields
are summarized as follows,
\brr \d A_i{}^\a &=& 2\bar \e_i{\z}^\a
     +2\rho^{\a\b}\ve_{ij}\bar\e^j {\z}_\b \,,\nonumber\\
     \d\zeta^\a &=& \Dslash A_i{}^\a\e^i
     +2X^0 A_i{}^{\a ({\rm z})} \ve^{ij}\e_j
     +2gX^\a{}_\b \, A_i{}^\b\ve^{ij}\e_j
     +A_i{}^\a\eta^i \,,
\label{hrules}\err
where $X^0$ is the scalar component of a background vector
multiplet which supplies the
gauge field for the central charge, and $X^\a{}_{\b}$ is
the scalar component of a Lie-algebra valued vector
multiplet\footnote{
   Our conventions are such that $X^\a_{\;\b}=X^I\,(t_I)^\a{}_{\;\b}$ and
   $\bar{X}^\a_{\;\b}=\bar{X}^I\,(t_I)^\a_{\;\b}$,
   where $t_I$ are the generators of the Lie algebra.
   Consistency requires that
   $(t_I)_\a^{\;\b} =-\rho_{\a\g}(t_I)^\g_{\;\eta}\rho^{\eta\b}$. A 
   nontrivial action requires the existence of an hermitean tensor 
   (which is not necessarily positive-definite, as one of the 
   hypermultiplets may act as a compensator), 
   which restricts the gauge group to a subgroup of (a noncompact 
   version of) USp($2r$). }. %
Central charge transformations commute with the
supersymmetry transformations when acting on the hypermultiplet
fields\footnote{%
   Later when we discuss the vector-tensor multiplet
   we will see that that $[\d_z(z), \d_Q (\e)]$ closes into the 
   tensor and vector gauge transformations that are associated to 
   the vector-tensor multiplet. The hypermultiplet is inert under 
   the latter transformations. }. %
It follows that the supersymmetry transformation for e.g. 
$A_i^{\;\a({\rm z})}$ is obtained from (\ref{hrules}) by placing a
``z'' index onto the rule for $A_i^{\;\a}$. Similarly,
the transformation rules for all fields higher in the hierarchy
can be obtained from those corresponding to lower-lying
fields by appending successive ``z'' indices. In order to close the
supersymmetry algebra an infinite number of constraints must be imposed.
The fields $(A_i^{\;\a},\zeta^\a,A_i^{\;\a({\rm z})})$ are not affected by the
constraints.
As a result they constitute the fundamental $8r+8r$ degrees of freedom
contained in the $r$ hypermultiplets. The constraints, which 
relate higher-$z$ elements
of the central charge hierarchy to the fundamental degrees of freedom,
are described by the following
two relationships,
\brr \zeta^{\a({\rm z})} &=& -\frac{1}{2\bar{X}^{0}}\Big[
     \rho^{\a\b}\Dslash\zeta_\b
     +\O^{0\,i}A_i^{\;\a({\rm z})}
     +g\O^{i\,\a}{}_{\!\b} A_i^{\;\b}
     +2g\bar{X}^\a{}_\b\zeta^\b
     +\ft18\s\cdot T_{ij}\varepsilon^{ij}\zeta^\a
     -\ft32\ve^{ij}\chi_iA_j^{\;\a} \Big] \,,\nonumber\\
     A_i^{\a({\rm zz})} &=& -\frac{1}{4|X^{0}|^2}\Big[
     (D^aD_a+\ft32D)A_i^{\;\a}
     +\ve_{ik}Y^{0\,jk}A_j^{\;\a({\rm z})}
     +2\Big(\rho^{\a\b}\bar{\O}_i^{0}\zeta_\b^{({\rm z})}
     -\ve_{ij}\bar{\O}^{0\,j}\zeta^{\a({\rm z})}\Big) \nonumber\\
     & & \hspace{.8in}
     +2g(\bar{X}^{0}X^\a{}_\b+X^{0}\bar{X}^\a{}_\b)A_i^{\;\b({\rm z})}
     +2g\Big(\rho^{\a\b}\bar{\O}_{i\b}{}^\g\,\zeta_\g
     -\ve_{ij}\bar{\O}^{j\,\a}{}_{\!\b}\zeta^\b \Big) \nonumber\\
     & & \hspace{.8in}
     +g\ve_{ik}Y^{jk\,\a}{}_\b\, A_j^{\;\b} 
     +2g^2\{\bar{X},X\}^\a{}_\b\, A_i^{\;\b}  \Big] \,.
\label{hconstraints}\err
All other constraints are obtained from these
by application of the central charge.

An important observation is that the constraints (\ref{hconstraints})
are algebraic relationships. For instance, the equation for
$\zeta^{\a({\rm z})}$ involves $\zeta_\b^{({\rm z})}$ on the right hand side,
through the covariant derivative
$D_\mu\zeta_\b=\der_\mu\zeta_\b-W_\mu^0\zeta_\b^{({\rm z})}+\cdots$.
Taking the complex conjugate of the equation for $\zeta^{\a({\rm z})}$, we
obtain the analogous equation for $\zeta_\a^{({\rm z})}$ which we may
then substitute back.  Similar manipulations may be done to
the equation for $A_i^{\a({\rm zz})}$. In this manner we can restructure
the constraints into the following form,
\brr \zeta^{\a({\rm z})}
     &=& -\ft12 X^0 \Big(\vert X^0\vert^2+\ft{1}{4}W_\mu^{0}W^{\mu 
    {0}}\Big)^{-1}\bpl  
     \rho^{\a\b}\,\dslash\zeta_\b+\cdots\bpr\,, \nonumber\\
     A_i^{\;\a({\rm zz})} &=& -\ft14 \Big(|X^{0}|^2  +\ft{1}{4} 
     W_\mu^{0}W^{\mu {0}}\Big)^{-1}\bpl
     \der^2 A_i^{\;\a} +\cdots\bpr \,.
\err

The above infinite-dimensional hierarchical structure of basic
units $(A_i{}^\a, \zeta^\a)$ endowed with 
an infinite sequence of constraints leaving precisely $8+8$ degrees of 
freedom, was worked out in \cite{DWLVP}. However, we should recall that 
this approach does not enable one to derive the most general 
couplings of hypermultiplets.
These can be obtained in the harmonic-superspace formulation, which avoids
the presence of an off-shell central charge at the expense of 
an infinite number of unconstrained fields. For the vector-tensor
multiplet there seems no way to avoid the central charge 
\cite{DK}. Therefore we use the same method as outlined above for the
construction of Lagrangians for interacting vector-tensor
multiplets. This is described in the next section. 
\vspace{.1in}

\subsection{The linear multiplet}
A linear multiplet contains three scalar fields transforming 
as an SU(2) triplet. The defining condition is that, under 
supersymmetry, these scalars transform into a doublet spinor. 
Furthermore it contains a Lorentz vector, subject to a 
constraint. The linear multiplet can transform in a real 
representation of some gauge group, as well as under a central 
charge. For this reason the supersymmetry transformations contain 
the Lie-algebra valued components of a vector multiplet 
associated with this gauge group. This is exactly the same as for 
the hypermultiplets, but here we do not introduce extra indices 
to indicate the matrix-valued character. The terms associated 
with the gauge group carry a coupling constant $g$. The 
central-charge transformations are simply incorporated into the 
generic gauge group and will not be indicated explicitly.
The Weyl and chiral weights and the fermion chirality of the
component fields of the linear multiplet are listed in table 2.2.

The transformation rules for the component fields of the linear multiplet
are as follows 
\brr \d L_{ij} &=& 2\bar{\e}_{(i}\vphi_{j)}
     +2\ve_{ik}\ve_{jl}\bar{\e}^{(k}\vphi^{l)}\,, \nonumber\\
     \d\vphi^i &=& \Dslash L^{ij}\e_j+\Eslash\ve^{ij}\e_j-G\e^i
     +2g\bar{X}L^{ij}\ve_{jk}\e^k + 2 L^{ij}\eta_j\,,\nonumber\\
     \d G &=& -2\bar{\e}_i\Dslash\vphi^i - \bar{\e}_i \bpl
     6 \chi_j L^{ij} + \ft12 \ve^{ij} \ve^{kl}\s \cdot T_{jk}
     \varphi_l \bpr \nonumber\\
     & &+2g\bar{X}\bpl\ve^{ij}\bar{\e}_i\vphi_j - {\rm h.c.} \bpr
     -2g\bar{\e}_i\O^j L^{ik}\ve_{jk} + 2 \bar \eta_i\varphi^i\,, 
    \nonumber\\
     \d E_a &=& 2\ve_{ij}\bar{\e}^i\s_{ab}D^b\vphi^j
     + \ft14 \bar {\e}^i \g_a \bpl 6 \ve_{ij} \chi_k L^{jk}
     - \ft12 \s\cdot T_{ij} \ve^{jk} \vphi_k\bpr \nonumber\\
     & &+2g\bar{X}\bar{\e}^i\g_a\vphi_i
     +g \bar{\e}^i\g_a\O^j L_{ij}  + \ft32 \bar 
     \eta^i\g_a\varphi^j\ve_{ij}+  {\rm h.c.}\,, 
\label{linear}
\err
where $L_{ij}=\ve_{ik}\ve_{jl}L^{kl}$ and
\bq   2D_a E^a= g \Big(\ft12 Y^{ij}L_{ij} 
      -2 X G-2\bar{\O}^i\vphi_i\Big) - 3 \bar {\vphi}^i \chi^j \ve_{ij} +
      {\rm h.c.} \,.
\eq
For $g=0$ the above constraint can be solved and $E^a$ can be 
written as the (supercovariant) field strength of a two-rank 
tensor gauge field $E_{\m\n}$. The solution takes the form
\bq
E^a =\ft12 i e^{-1} e^a_\m \ve^{\m\n\rho\s} D_\n E_{\rho\s}\,.
\eq
The resulting multiplet is known as the $N=2$ tensor multiplet.

\begin{figure}
\begin{center}
\begin{tabular}{|c||cccc||ccc||cccc|}
\hline
&
   \multicolumn{4}{c||}{vector multiplet}  &
   \multicolumn{3}{c||}{hypermultiplet}    &
   \multicolumn{4}{c|}{linear multiplet}   \\
\hline
\hline
field          &
   $X^I$         &
   $\O_i^{\,I}$      &
   $W_\mu^{\,I}$     &
   $Y_{ij}^{\,I}$    &
   $A_i^\a$    &
   $\zeta^\a$  &
   $A_i^{\a({\rm z})}$   &
   $L_{ij}$    &
   $\varphi^i$ &
   $G$         &
   $E_a$       \\[.5mm]
\hline
\hline
w         & $1$ & $\ft32$  & $0$ & $2$ &
            $1$ & $\ft32$  & $1$ &
            $2$ & $\ft52$  & $3$ & $3$ \\[.5mm]
\hline
c         & $-1$& $-\ft12$ & $0$ & $0$ &
            $0$ & $-\ft12$ & $0$ &
            $0$ & $\ft12$  & $1$ & $0$ \\[.5mm]
\hline
$\gamma_5$&     &  $+$     &     &     &
                &  $-$     &     &
                &  $+$     &     &     \\[.5mm]
\hline
\end{tabular}\\[.13in]
\parbox{4.6in}{Table 2.2: Weyl and chiral weights ($w$ and $c$, respectively)
             and fermion chirality ($\gamma_5$)
             of the vector, hyper and linear multiplet component fields.}
\end{center}
\end{figure}

\subsection{Multiplet calculus}
The identification of the various rules for multiplying multiplets 
is a central aspect of the multiplet calculus. This has been 
explicitly described in previous papers \cite{DWVHVP,DWLVP}. 
There are product rules that define how to construct multiplets 
from products of certain other multiplets. For some of the 
multiplets one can find density formulae, which yield a 
superconformally invariant action upon integration over 
spacetime. In the context of our work here the most relevant 
density formula is the one involving an abelian vector and a linear 
multiplet. The linear multiplet can transform under a central 
charge, in which case the vector multiplet must be the one that 
supplies the gauge field for the central charge transformations. 
Apart from this the linear multiplet must be neutral under the 
gauge group. The density formula reads, 
\brr
e^{-1}{\cal L}&=& X^0 G - \bpl \ft14 Y^{0\,ij} + \ft12 \bar
\psi^{i}_\m  \g^\m 
\O^{0\,j} + \bar X^0 \bar \psi^i_{\mu} \s^{\mu\nu} \psi_\nu^j\bpr L_{ij}
+ \bar\vphi^i \bpl \O_i^0 + X^0 \g^\m \psi_{\m i} \bpr\nonumber\\
& &- \ft12 W_a^0 \bpl E^a + 2 \bar \vphi^i \s^{ab} \psi_b^j 
\ve_{ij} - \ft12 \ve^{abcd}\, \bar \psi_{b\, k}\g_c\psi_d^i L_{ij}
\ve^{jk} \bpr + {\rm h.c. }\,, 
\label{linaction}
\err
where the vector-multiplet fields carry a superscript ``0" to 
indicate that they belong to an abelian vector multiplet, possibly 
associated with central-charge transformations. 
\setcounter{equation}{0}
\section{The vector-tensor multiplets}

\subsection{Central charges and Chern-Simons terms}
{}From off-shell counting it follows immediately that the vector-tensor
multiplet must be subject to a central charge when it is based on a finite 
number of off-shell components. Just as in \cite{vt1,vt2} we use the
same strategy as presented for hypermultiplets in the previous 
section.
The basic unit of the vector-tensor multiplet consists of a 
scalar field $\phi$, a vector gauge
field $V_\mu$, a tensor gauge field $B_{\mu\nu}$ and a doublet of
spinors $\l_i$. This unit consists of seven bosonic and eight
fermionic components. To close the supersymmetry algebra off
shell, we must 
assume the existence of an infinite hierarchy of these units, again
distinguished by appending successive indices ``z''. The central
charge then raises the number of ``z'' indices, such as, for instance,
in $\d_{z}\,\phi=z\phi^{({\rm z})}$. Successive applications thus 
generate a sequence of terms,
\bq \phi\longrightarrow\phi^{({\rm z})}\longrightarrow
\phi^{({\rm zz})}   \longrightarrow {\rm etcetera}\,,
\label{hierarchy}
\eq
and similarly on all other fields. It will turn out that $\phi^{({\rm z})}$
corresponds to an auxiliary field. All 
other objects in the hierarchy, $\phi^{({\rm zz})}$, $V_{\mu}^{({\rm z})}$,
$V_\mu^{({\rm zz})}$, etcetera, are dependent, and will be given by particular
combinations of the independent fields. Hence we end up with precisely
$8+8$ degrees of freedom. 

In order to couple the vector-tensor multiplet to supergravity we 
employ the superconformal multiplet calculus. When the supersymmetry
is local then also the central-charge transformations must be local.
Therefore we must couple the vector-tensor multiplets to at least one
vector multiplet, whose gauge field couples to the central
charge. However, for reasons that have been described in \cite{vt2}, it
is advisable to couple the vector-tensor multiplet to a more
general background of vector multiplets, so we consider $n$ vector
multiplets. One of these provides the gauge field for the central charge,
which we denote by $W_\mu^0$. This must be an abelian gauge field. The
remaining $n-1$ vector multiplets supply additional background gauge fields
$W_\mu^A$, which need not be abelian. The index $A$ is taken to run from
$2$ to $n$, for reasons we explain shortly. Also, since $W_\mu^0$ 
is the gauge field for the central  
charge, the associated transformation parameter $\t^0$ is identified with
the central charge parameter $z$ introduced above, i.e., 
$z\equiv\t^0$.  The vector gauge transformations act as follows 
on the background gauge fields, 
\bq 
    \d W_\mu^0=\der_\mu z \,,\qquad
    \d W_\mu^A=\der_\mu\t^A+f^A{}_{\!BC}\t^BW_\mu^C\,. 
\eq

In addition to the central charge, the vector-tensor multiplet
has its own gauge transformations associated with the tensor 
$B_{\mu\nu}$ and the vector $V_\mu$.
We reserved the index $1$ for the vector field $V_\mu$ of the
vector-tensor multiplet. (The reason for this choice is based on the dual
description of our theory, where the vector-tensor multiplet is replaced
with a vector multiplet, so that the dual theory involves $n+1$ vector
multiplets.) 
In the interacting theory, the tensor field
$B_{\mu\nu}$ necessarily couples to Chern-Simons forms. This coupling is
evidenced by the transformation behavior of the tensor. To illustrate
this, if we ignore the central charge (other than its contribution to
$W^0_\m$), then the vector field of the vector-tensor multiplet would
transform as
\bq \d V_\mu=\der_\mu\t^1 \,, \label{simv}
\eq
and the tensor field would transform as
\bq \d B_{\mu\nu}=2\der_{[\mu}\Lambda_{\nu]}
    +\eta_{IJ}\,\theta^I\der_{[\mu}W_{\nu]}^J,
\label{simb}
\eq
where $\t^I$ and $\Lambda_\mu$ are the parameters of the transformations
gauged by $W_\mu^I$ and $B_{\mu\nu}$ respectively, and the index $I$ is
summed from $0$ to $n$. As mentioned above, in this context $W_\mu^1$ is
identified with $V_\mu$. Closure of the combined vector and tensor gauge
transformations requires that $\eta_{IJ}$ be a constant tensor
invariant under the gauge group. There is an ambiguity in the
structure of $\eta_{IJ}$, which derives from the possibility of performing
field redefinitions.  Without loss of generality, $\eta_{IJ}$
can be modified by absorbing a term proportional to
$W_\mu^IW_\nu^J$ times some group-invariant antisymmetric tensor 
into the definition of the tensor field $B_{\mu\nu}$. 
Without loss of generality, we thus remove all components of
$\eta_{IJ}$ except for $\eta_{11}, \eta_{1A}$ and $\eta_{AB}$,
and also we 
render $\eta_{AB}$ symmetric. Also note that, since $\eta_{1A}$ 
is invariant under the gauge group, it follows that 
$\eta_{1A}W_\mu^A$ is an abelian gauge field.

The situation is actually more complicated, since $V_\mu$ and
$B_{\mu\nu}$ are also subject to the central-charge
transformation. As described above, under this transformation
these fields transform into complicated expressions, denoted 
$V_\mu^{({\rm z})}$ and $B_{\mu\nu}^{({\rm z})}$, respectively, 
which involve other fields of the theory. Accordingly, we deform the
transformation rule (\ref{simv}) to
\bq \d V_\mu = \der_\mu \t^1+z V_\mu^{({\rm z})} \, ,
\eq
and, at the same time, (\ref{simb}) to
\bq \d B_{\mu\nu} = 2\der_{[\mu}\Lambda_{\nu]}
    +\eta_{11}\,\theta^1\der_{[\mu}V_{\nu]}
    +\eta_{1A}\,\theta^1\der_{[\mu}W_{\nu]}^A
    +\eta_{AB}\,\theta^A\der_{[\mu}W_{\nu]}^B
    +z B_{\mu\nu}^{({\rm z})} \,.
\eq
All $\t^0$-dependent terms, including any such Chern-Simons contributions,
are now contained in $V_\m^{({\rm z})}$ and $B_{\mu\nu}^{({\rm 
z})}$, which are determined by closure of the full algebra, 
including supersymmetry. The deformed transformation rules must 
still lead to a closed gauge algebra. In particular one finds that
\bq [\d_z ( z), \d_{{\rm vector}} (\theta^1)] =
    \d_{{\rm tensor}} (\ft12z\, \h_{11}\,\theta^1\, V_\mu^{({\rm z})}) \, .
\eq
This implies that $V_\mu^{({\rm z})}$ and the combination
$B_{\mu\nu}^{({\rm z})}+\eta_{11}V_{[\mu}V_{\nu]}^{({\rm z})}$ 
both transform covariantly under the central charge, but are 
invariant under all other gauge symmetries. However, under local 
supersymmetry, they do not transform covariantly, as we will see 
below (cf. \ref{VzBz}). The resulting   
gauge algebra now consists of the standard 
gauge algebra for the vector fields augmented by a tensor gauge
transformation.
Observe that we have neither specified $V^{({\rm z})}_\m$ nor 
$B_{\mu\nu}^{({\rm z})}$,
which are determined by supersymmetry and will be discussed in the next
section. As it turns out these terms give rise to additional 
Chern-Simons terms involving $W_\m^0$ that depend on the scalar 
fields. The presence of these terms is a direct result of the 
deformation of the standard algebra of tensor and vector gauge 
transformations. 

Before giving specific results on the local supersymmetry 
transformations, we discuss a crucial feature of our results. It 
turns out \cite{vt1,vt2} that the coefficients $\eta_{IJ}$ that encode 
the Chern-Simons terms cannot all be zero, as otherwise the 
supersymmetry variations turn singular and supersymmetric 
completions in the action will vanish. In fact, one can show that 
there are just two inequivalent representations of the 
vector-tensor multiplet. One is the case where $\eta_{11}=0$. In 
this case there is no Chern-Simons coupling between the tensor 
and the vector fields of the vector-tensor multiplet. The choice 
$\eta_{11}=0$ removes the conspicuous self-interaction between 
the vector-tensor multiplet fields and in fact  
the supersymmetry transformations become linear in these fields 
(but not in the background fields) and the action quadratic. 
However, in this case not all the $\eta_{1A}$ Chern-Simons 
coefficients can vanish simultaneously. Therefore we are 
dealing with at least three abelian gauge fields, namely, 
$W_\m^0$, $\eta_{1A}W^A_\m$ and $V_\m$. In the case of rigid 
supersymmetry, one can freeze some or all of the vector multiplets to a 
constant, but this will not alter the structure of the couplings. 

This first class seems to coincide with the theories one obtains 
by reducing (1,0) tensor multiplets in six spacetime dimensions 
to four dimensions. The tensor multiplet comprises a scalar, a 
self-dual tensor gauge field and a symplectic Majorana spinor. 
The self-dual tensor field decomposes in four dimensions into the 
vector and tensor gauge fields of the vector-tensor multiplet. To 
have also a vector field that couples to the central charge 
presumably requires the dimensional reduction of a theory of 
tensor multiplets coupled to supergravity. A recent study of various Chern-Simons 
terms in six dimensions was carried out in \cite{BSS}. 

The second, inequivalent class of couplings is characterized by 
the fact that $\eta_{11}\not=0$. In that case, it turns out that 
one can absorb certain terms of the background multiplets into 
the definition of the vector-tensor fields such that all the 
coefficients $\eta_{1A}$ vanish \cite{vt2}. In this case we have 
at least two abelian vector fields, namely $W^0_\m$ and $V_\mu$. 

Hence in practical situations the Chern-Simons coefficients can 
be restricted to satisfy either $\eta_{11}=0$ or $\eta_{1A}=0$. In the 
following we will not pay much attention to this fact, but simply 
evaluate the transformation rules and the action for general 
values of the coefficients $\eta_{11}$, $\eta_{1A}$, $\eta_{AB}$.

\subsection{The vector-tensor transformation rules\label{s:vttrans}}
In \cite{vt1,vt2} the transformation rules for the vector-tensor
multiplet have been determined by imposing the supersymmetry
algebra iteratively on the multiplet component fields. In this
procedure, the supersymmetry
transformation rules for vector multiplets remain unchanged. Therefore, the
algebra represented by the vector-tensor multiplet in the presence of a
vector multiplet background is fixed up to gauge transformations which
pertain exclusively to the vector-tensor multiplet. The most relevant
commutator in this algebra involves two supersymmetry 
transformations and was given in (\ref{qqcomb}). In the 
present situation we have that 
\brr \d_{\rm gauge} &=&
     \d_z\bpl 4\ve^{ij}\bar{\e}_{2i}\e_{1j}X^0+ {\rm h.c.} \bpr
     +\d_{{\theta^A}} \bpl 4\ve^{ij}\bar{\e}_{2i}\e_{1j}X^A+ {\rm h.c.} \bpr
     \nonumber\\
     &&+\d_{\rm vector}\bpl \t^1(\e_1,\e_2)\bpr
     +\d_{\rm tensor}\bpl \Lambda_\mu(\e_1,\e_2)\bpr \,.
\label{algebra}
\err
The field $X^0$ is the complex scalar of the vector multiplet
associated with the central charge\footnote{%
   Henceforth we will suppress the superscript on $X^0$ and define
   $X\equiv X^0$ to simplify the formulae.}.
The field-dependent parameters  $\t^1(\e_1,\e_2)$ and 
$\Lambda_\mu(\e_1,\e_2)$ are found by imposing the 
$Q$-super\-symmetry commutator on the vector-tensor multiplet. They 
will be specified in due course. 

In this paper we repeat the derivation of the transformation 
rules, but now in the context of local supersymmetry. This means 
that we follow the same procedure, but now in a background of conformal 
supergravity combined with vector multiplets. Because the transformation 
rules for the superconformal  
fields are also completely known, the supersymmetry 
algebra is determined up to the gauge  and central-charge transformations 
associated with the vector-tensor multiplet itself. 
The procedure followed in \cite{vt1,vt2} is tailor-made for an 
extension to local supersymmetry. First of all, we already  
insisted on rigid scale and chiral invariance. Because of that,  
the scalar fields of the vector multiplets will play the role of 
compensating fields to balance possible differences in scaling 
weigths of the various terms. Secondly, one of the vector 
multiplets was required to realize the central charge in a local 
fashion. In the context of the superconformal multiplet calculus, 
local dilations, chiral and central-charge transformations are 
necessary prerequisites for the coupling to supergravity. 

As was already discussed in \cite{vt1}, there remains some 
flexibility in the assignment of the scaling and chiral weights 
for the vector-tensor multiplet. By exploiting the scalar
fields of the vector multiplets we may arbitrarily adjust the 
weights for each of the vector-tensor 
components by suitably absorbing functions of $X$ and $X^A$. In 
this way we choose the weights for the vector-tensor components 
to be as shown in table 3.1.
\begin{figure}
\begin{center}
\begin{tabular}{|c||ccccc|}
\hline
& \multicolumn{5}{c|}{vector-tensor multiplet} \\
\hline
\hline
field & $\phi$  & $V_\mu$ & $B_{\mu\nu}$ & $\l_i$ & $\phi^{({\rm z})}$ \\[.5mm]
\hline
\hline
$w$ & $0$ & $0$ & $0$ & $\ft12$ & $0$  \\[.5mm]
\hline
$c$ & $0$ & $0$ & $0$ & $\ft12$ & $0$  \\[.5mm]
\hline
$\gamma_5$& & & & $+$ &  \\[.5mm]
\hline
\end{tabular}\\[.13in]
\parbox{5in}{Table 3.1:
   Scaling and chiral weights ($w$ and $c$, respectively) and fermion
   chirality ($\gamma_5$) of the vector-tensor component fields.}
\end{center}
\end{figure}
The bosonic vector-tensor fields must all have chiral weight $c=0$ since
they are all real. To avoid a conflict between scale transformations and
vector-tensor gauge transformations we adjusted $V_\mu$ and $B_{\mu\nu}$
to be also neutral under scale transformations. Note that there 
remains a freedom to absorb additional 
combinations of the background fields into the definition of $\phi$ and
$\l_i$. Furthermore, the fields $V_\m$ and $B_{\mu\nu}$ can be redefined by
appropriate additive terms. Needless to say, it is important to 
separate relevant terms in the 
transformation rules from those that can be absorbed into such field
redefinitions. In deriving our results this aspect has received proper
attention.  

In order to define the vector-tensor multiplet as a 
superconformal multiplet, we must also choose the assignments under 
the special $S$-supersymmetry transformations (which in turn determine 
the behaviour under special conformal boosts $K$). We have 
assumed that the scalar $\phi$ is $S$- and $K$-invariant, which 
leads to consistent results. While 
this is a natural assignment for the lowest-dimensional component 
of a supermultiplet, we found no rigorous arguments to rule out other 
assignments. The choice we made is the simplest one and, as it 
turns out, implies 
that all the vector-tensor fields remain $S$- and $K$-invariant.  
The latter follows from the commutator of $Q$- with $S$-supersymmetry, 
and subsequently, by using the $[S,S]$ commutation relation, 
which yields a $K$-transformation.

The transformation rules coincide with the ones found in 
\cite{vt1,vt2} apart from the presence of certain 
covariantizations. As before we suppress nonabelian terms for the 
sake of clarity; they are 
not important for the rest of this paper. We are not aware of arguments
that would prevent us from switching on the nonabelian 
interactions. Furthermore we introduce the following notation 
for homogeneous, holomorphic functions of zero degree that occur 
frequently in our equations, 
\bq g=i\eta_{1A}\frac{X^A}{X}\,, \qquad
    b=-\ft14 i\eta_{AB}\frac{X^AX^B}{X^2} \,.
\label{bgdef}
\eq
For arbitrary Chern-Simons coefficients $\eta_{IJ}$, the 
transformation rules under $Q$-supersymmetry are (we 
emphasize that in the remainder of this section and in section~4, 
the index $I$ does not take the value $I=1$),  
\brr \d\phi &=& \bar{\e}^i\l_i +\bar{\e}_i\l^i \,,\nonumber\\
     \d V_\mu &=& i\ve^{ij}\bar{\e}_i\g_\mu \bpl 2X\l_j+\phi\O_j^0\bpr
     -iW_\mu^0\,\bar{\e}^i\l_i + 2i\phi X \ve^{ij} \bar{\e}_i \psi_{\mu j}
     + {\rm h.c.}  \,, \nonumber\\
     \d B_{\mu\nu} &=& -2\bar{\e}^i\s_{\mu\nu}|X|^2\bpl 4\eta_{11}\phi
     - 2 {\rm Re}\, g \bpr\l_i  \nonumber\\
     & & -2\bar{\e}^i\s_{\mu\nu}\bar{X}\bpl 2\eta_{11}\phi^2\O_i^0
     +\phi\bar{X}\der_{\bar{I}}\bar{g}\,\O_i^I
     -4i{\rm Re}[\der_I(Xb)]\O_i^I \bpr \nonumber\\
     & & -2 \bar\e^i\g_{[\mu}\psi_{\nu]i} \bar X\bpl 2\eta_{11}\phi^2 X
     +\phi\bar X \der_{\bar I}\bar g X^I -4i{\rm Re}[\der_I(Xb)] X^I\bpr
      \nonumber\\
     & & +i\ve^{ij}\bar{\e}_i\g_{[ \mu}V_{\nu ]} \bpl \eta_{11}
     ( 2X\l_j +\phi\O_j^0 ) -i \eta_{1A}\O_j^A\bpr \nonumber\\
     & & +2i \ve^{ij}\bar{\e}_i\psi_{j[\mu }V_{\nu ]} X
     \bpl \eta_{11} \phi - g \bpr  \nonumber\\
     & & +\ve^{ij}\bar{\e}_i\g_{[\mu}W_{\nu]}^0\bpl 2X
     (2\eta_{11}\phi-g)\l_j +\eta_{11}\phi^2\O_j^0 -i\eta_{1A}\phi\O_j^A
     -4i\der_I(Xb)\O_j^I\bpr  \nonumber\\
     & & + 2 \ve^{ij}\bar{\e}_i\psi_{j[\mu}W_{\nu]}^0 X
     \bpl\h_{11}\phi^2 -\phi g -4i b  \bpr  \nonumber\\
     & & + \ve^{ij}\bar{\e}_i \g_{[ \mu}W_{\nu ]}^A \h_{AB} \O_j^B
     + 2\ve^{ij}\bar{\e}_i \psi_{j[\mu } W_{\nu ]}^A  \h_{AB} X^B
     \nonumber\\
     & & -i\eta_{11}W_{[\mu}^0V_{\nu]}\bar{\e}^i\l_i + {\rm h.c.} \,,
     \nonumber\\
     \d\l_i &=&
     \bpl\Dslash\phi-i \hatVslash^{({\rm z})}\bpr\e_i
     -\frac{i}{2X}\ve_{ij}\s\cdot\bpl \F^-(V) -i\phi\F^{- 0}\bpr\e^j
     +2\ve_{ij}\bar{X}\phi^{({\rm z})}\e^j \nonumber\\
     & & -\frac{1}{X}(\bar{\e}^j\l_j)\O_i^0
     -\frac{1}{X}(\bar{\e}^j\O_j^0)\l_i \nonumber\\
     & & -\frac{1}{2X (2\eta_{11}\phi - {\rm Re}\, g )} \e^j 
\Big[
     2\eta_{11}\phi^2Y_{ij}^0 +\phi\bar{X}\der_{\bar{I}}\bar{g}\,Y_{ij}^I
     -4i{\rm Re}\,\der_I(Xb)Y_{ij}^I\nonumber\\
     & & \hspace{1.6in} -2\eta_{11} \bpl X\bar{\l}_i\l_j
     -\bar{X}\ve_{ik}\ve_{jl}\bar{\l}^k\l^l\bpr \nonumber\\
     & & \hspace{1.6in} +X \bpl X\der_I g \,\bar{\O}_{(i}^I\l_{j)}
     -\bar{X}\ve_{ik}\ve_{jl}\der_{\bar{I}}\bar{g}\,
     \bar{\O}^{I(k}\l^{l)}\bpr \nonumber\\
     & & \hspace{1.6in}
     + i \bpl \der_I\der_J(Xb)\,\bar{\O}_i^I\O_j^J
     + \ve_{ik}\ve_{jl}\,\der_{\bar{I}}\der_{\bar{J}}(\bar{X}\bar{b})
     \bar{\O}^{Ik}\O^{Jl}\bpr\Big]\,.\;
\label{momrules}
\err
Except from the explicit gravitino fields in the variations of 
$V_\m$ and $B_{\m\n}$, all extra covariantizations are implicitly 
contained in covariant derivatives and field strengths. 

Let us now first define a number of quantities that appear in 
(\ref{momrules}) or are related to them. The supercovariant field 
strengths for the vector-tensor multiplet gauge fields are equal 
to 
\brr \label{fs}
    \F_{\mu\nu} (V) &=& 2\der_{[\mu}V_{\nu]} -2W^0_{[\mu}V_{\nu]}^{({\rm z})}
     + \ft14i\phi\Big[\bar{X} T_{\mu\nu}^{ij} \ve_{ij}-{\rm 
h.c.}\Big]  \nonumber\\
     && - i\Big[ \ve^{ij} \bar{\psi}_{i[\mu} \g_{\nu ]}\bpl 2X\l_j
     +\phi\O_j^0\bpr + \phi X \ve^{ij} \bar{\psi}_{\mu i} \psi_{\nu j}
     - {\rm h.c.}\Big] \,,
     \nonumber\\
     H^\mu &=& \ft12 {i} e^{-1} \ve^{\mu\nu\l\s} \Big[\der_\nu B_{\l\s}
     - \eta_{11} V_\nu \,\der_\l V_\s -\eta_{1A} V_\nu \,\der_\l W_\s^A
     \nonumber\\
     && \hspace{1.8cm} - \eta_{AB} W_\nu^A  \der_\l W_\s^B - W_\nu^0
     \bpl B_{\l\s}^{({\rm z})} + \h_{11} V_\l V_\s^{({\rm z})} 
     \bpr \Big] \\
     &&-\Big[ i {\bar\psi}^i_\nu \s^{\m\n}\Big( 2 |X|^2 \bpl 2\eta_{11}\phi
     - {\rm Re}\, g \bpr\l_i \nonumber\\
     &&\hspace{2cm} + \bar{X}\bpl
     2\eta_{11}\phi^2\O_i^0 +\phi\bar{X}\der_{\bar{I}}\bar{g}\,\O_i^I
     -4i{\rm Re}[\der_I(Xb)]\O_i^I \bpr\Big) + {\rm h.c.}\Big]  \nonumber\\
     && + \ft14 {i} e^{-1} 
    \ve^{\mu\nu\l\s}{\bar\psi}^i_\nu \g_\l \psi_{\s i} 
    \Big[
     \bar X\bpl 2\eta_{11}\phi^2 X +\phi\bar X X^I\,\der_{\bar I}\bar g 
     -4iX^I\,{\rm Re}[\der_I(Xb)] \bpr +{\rm h.c.} \Big] \,
    .\nonumber 
\err
The Bianchi identities corresponding to the field strengths (\ref{fs}) are
straightforward to determine and read,
\brr 
&& D_\mu\Big(\tilde{\F}^{\mu\nu}(V) + \ft14i\phi (\bar XT^{\m\n\,
  ij}\ve_{ij} +  X T_{ij}^{\m\n}\ve^{ij})  \Big)  
\nonumber\\
&&\hspace{1.15cm}   =-V_\mu^{({\rm z})}\Big[\tilde{\F}^{0\mu\nu} 
-\ft14 (\bar XT^{ij\, \m\n}\ve_{ij} -  X T_{ij}^{\m\n}\ve^{ij}) 
\Big]  - \ft34 
   i\Big[\varepsilon_{ij} \bar \chi^i\g^\n (2\bar X \l^j+ 
   \phi\O^{j0})+ {\rm h.c.} \Big]  \,,\nonumber  \\
&&     D_\mu H^\mu =  - \ft14 i\Big[\eta_{11}\, \tilde{\cal F}_{\m\n}(V)\,
      {\cal F}^{\m\n}(V) +\eta_{1A}\,\tilde {\cal F}_{\m\n}(V)\,
      {\cal F}^{\m\n A} +\eta_{AB}\, \tilde{\cal F}_{\m\n}^{A}\,
      {\cal F}^{\m\n B} + 2 \tilde \F^{0}_{\m\n}\hat B^{\m\n\,({\rm 
   z})}\Big] 
      \nonumber\\    
     &&\hspace{1.6cm} -\ft1{16}i\Big[T_{ij}^{\m\n}\Big(
       2\eta_{11}\, \phi X\,{\cal F}_{\m\n}(V) +\eta_{1A}(X^A \,
  {\cal F}_{\m\n}(V) +i \phi X {\cal F}_{\m\n}^A) +2\eta_{AB}\, 
  X^A\,{\cal F}_{\m\n}^B \nonumber \\
    && \hspace{3.4cm}   + 2X \hat B^{({\rm z})}_{\m\n}+X  \, 
\F^{0}_{\m\n} (\eta_{11} 
\phi^2 -\phi\,g -4ib)\Big)  - {\rm h.c.}\Big] \nonumber   \\
     &&\hspace{1.6cm} + 3 i( \bar \l_i\chi^i- \bar\l^i\chi_i) \,
   \vert X\vert^2  
    (2\eta_{11} \phi -  {\rm Re}(g)) \nonumber \\
      &&\hspace{1.6cm} - \ft32 i\Big[X \,\bar \chi_i (2\eta_{11} \phi^2 
\O^{i0}  + \phi X \der_{ I} g \O^{Ii} + 4i {\rm Re}[\pa_I(Xb)] 
\O^{Ii}) - {\rm h.c.} \Big] \,. 
\label{bianchis}
\err
Observe that the Bianchi identity for $H_\mu$ is not linear in the
vector-tensor fields. On the right-hand side there are nonlinear terms that
are either of second-order (the term proportional to $\eta_{11}$) or of
zeroth-order (the term proportional to $\eta_{AB}$) in the vector-tensor
fields. Furthermore the quantity $\hat B_{\mu\nu}^{({\rm z})}$ does not depend
homogeneously on the vector-tensor fields either as will become clear soon.
Hence, generically the vector-tensor multiplet is realized in a nonlinear
fashion, as we have already pointed out in the previous 
subsection. 

Furthermore, the following quantities appear in the above 
formulae, which are the supercovariant part of the 
$z$-transformed vector and tensor fields,  
\brr {\hat V}_a^{({\rm z})} &=& \frac{-1}{2|X|^2(2\eta_{11}\phi- 
   {\rm Re}\,g) } 
     \bl H_a - \Big[ i X D_a \bar{X}^I \bpl
     2\eta_{11}\phi^2 \d_I{}^0 +\phi\bar{X}\der_{\bar{I}}\bar{g}
     -4i{\rm Re}[\der_I(Xb)] \bpr + {\rm h.c.} \Big]\br \nonumber\\
     & & + {\rm fermion\ terms}\,, \phantom{\Big[ } \nonumber\\
     {\hat B}_{ab}^{({\rm z})} &=& -\ft12 {\rm Im}\,g \, \F_{ab}(V)
     +\ft12{i} (2\eta_{11}\phi- {\rm Re\,}g)\tilde{\F}_{ab}(V)
     -\ft12\phi(\eta_{11}\phi-{\rm Re}\,g)\F_{ab}^0 \nonumber\\
     & & +\ft12i  \phi{\rm Im}(X\der_I g)\tilde{\F}_{ab}^I
     +4{\rm Im}\Big[\der_I(Xb)\F_{ab}^{I-} \Big]
     + {\rm fermion\ terms} \,.
\label{vzbz}
\err
The caret indicates that these expressions are fully covariant 
with respect to all local symmetries; they do not coincide with 
the image of $V_\m$ and $B_{\m\n}$ under the central charge, 
$V_\m^{({\rm z})}$ and $B_{\m\n}^{({\rm z})}$. The latter are given by 
\brr \label{VzBz}
  V_\mu^{({\rm z})} &=& {e_\mu}^a {\hat V}_a^{({\rm z})} 
    + \ft12\bpl i \bar{\psi}^i_\mu \l_i + {\rm h.c.} \bpr \,,
\nonumber\\
   B_{\mu\nu}^{({\rm z})}&=& {e_\mu}^{[a}{e_\nu}^{b]} {\hat 
  B}^{({\rm z})}_{ab}  
     - \h_{11} V_{[\mu} V^{({\rm z})}_{\nu ]} \nonumber\\
     &&+\ft12 \Big[X \ve^{ij} ( \bar{\psi}_{\mu i} \psi_{\nu j}
     +\ft{1}{4}T_{\mu\nu\, ij})(\h_{11}\phi^2-\phi g-4ib )
     +  2 X\ve^{ij} \bar{\psi}_{i[\mu} \g_{\nu ]}\l_j \,(2\eta_{11}\phi-g)
      \nonumber\\
     &&\hspace{8mm} +  \ve^{ij} \bar{\psi}_{i[\mu} \g_{\nu ]}
     \bpl \eta_{11}\phi^2\O_j^0
     -i\eta_{1A}\phi\O_j^A -4i\der_I(Xb)\O_j^I\bpr + {\rm 
h.c.}\Big] \,.  
\err
There are of course similar expressions for $\l_i^{({\rm z})}$ and 
$\phi^{({\rm zz})}$, which are of less direct relevance. Because 
the fields $\phi$ and $\l_i$ are themselves covariant, the action of 
the central charge will yield covariant expressions. 
 
The results for the central charge transformations are determined 
from the commutator, 
\bq [\d_Q(\e), \d_z( z)] = \d_{\rm vector}\bpl i z \bar{\e}^i \l_i
     + {\rm h.c.} \bpr + \d_{\rm tensor}\bpl\Lambda_\mu (\e, z) 
\bpr\,,
\eq
where     
\brr     
    \Lambda_\mu (\e , z) &=&  \ft{1}{2} z \ve^{ij} \bar{\e}_i \g_\mu
     \bpl 2X(2\eta_{11}\phi-g)\l_j +\eta_{11}\phi^2\O_j^0
     -i\eta_{1A}\phi\O_j^A -4i\der_I(Xb)\O_j^I\bpr  \nonumber\\
     &&+z \ve^{ij} \bar{\e}_i \psi_{\mu j} X \bpl\h_{11}\phi^2 -\phi g -4ib
     \bpr +\ft12{i} z \h_{11} V_\mu \bar{\e}^i \l_i + {\rm 
h.c.}\,,
\err
which implies that the supersymmetry transformations of 
$\phi^{({\rm z})}$, $\l_i^{({\rm z})}$ are just the 
$z$-transformed versions of $\d_Q  
\phi, \d_Q \l_i$ as given in (\ref{momrules}). 
Hence, with the exception of $\phi^{({\rm z})}$ all the 
$z$-transformed fields are subject to constraints. By acting on 
these constraints with central-charge
transformations, one recovers an infinite hierarchy of constraints.  These
relate the components of the higher multiplets $(V_\mu^{({\rm z})},
B_{\mu\nu}^{({\rm z})}, \l_i^{({\rm z})}, \phi^{({\rm zz})})$, 
etcetera to the lower ones, in 
such a way as to retain precisely $8+8$ independent degrees of freedom.
 
At this point we specify the expressions for the vector and 
tensor gauge transformations in the commutator \eqn{algebra}, 
\brr \theta^1 (\e_1, \e_2) &=& 4i\phi X \,\ve^{ij} \bar{\e}_{i2} 
\e_{j1} + {\rm h.c.} \,,
     \nonumber\\
     \Lambda_\mu (\e_1, \e_2) &=&   2 \bar{\e}^i_2 \g_\mu 
\e_{i1}\,
     \bar X\bpl 2\eta_{11}\phi^2 X +\phi\bar X X^I\der_{\bar I}\bar g
     -4iX^I{\rm Re}[\der_I(Xb)]\bpr \nonumber\\
     && + 2i \ve^{ij} \bar{\e}_{i2} \e_{j1} \,X \bpl V_\mu (\h_{11}\phi - g)
     -i W_\mu^0 (\h_{11}\phi^2 -\phi g -4ib) \bpr  \nonumber\\
     && + 2 \ve^{ij} \bar{\e}_{i2} \e_{j1} \,W_\mu^A \h_{AB} X^B 
+ {\rm h.c.}\,. 
\err

We close this section with a number of supersymmetry 
variations of various quantities defined above. 
The supercovariant field strengths transform as follows:
\brr \d \F_{ab} (V) &=& -2i\ve^{ij} \bar{\e}_i\g_{[a}D_{b]}
     \bpl 2X\l_j + \phi \O^0_j\bpr -2 \ve^{ij} 
     \bar{\e}_i\g_{[a}\O^0_j\, 
     {\hat V}_{b]}^{({\rm z})} -i\bar{\e}^i\l_i \F_{ab}^0 \nonumber\\
     &&-2 i \ve^{ij} \bar{\h}_i \s_{ab} \bpl 2X \l_j + \phi \O^0_j \bpr
     + {\rm h.c.} \,,\nonumber\\
     \d H^a &=& 4i \bar{\e}^i\s^{ab}D_b \Big[
     |X|^2 \bpl 2\eta_{11}\phi -  {\rm Re}\, g\bpr\l_i\Big] 
     \nonumber\\ 
     && + 2i \bar{\e}^i\s^{ab}D_b \Big[ \bar{X}\bpl 2\eta_{11}\phi^2\O_i^0
     +\phi\bar{X}\der_{\bar{I}}\bar{g}\,\O_i^I
     -4i{\rm Re}[\der_I(Xb)]\O_i^I \bpr \Big]  \nonumber\\
     &&+\ft{3}{2} i \bar{\e}^i\g^a\chi_i \,\bar X\bpl
     2\eta_{11}\phi^2 X +\phi\bar X \der_{\bar I}\bar g X^I
     -4i{\rm Re}[\der_I(Xb)] X^I\bpr  \nonumber\\
     &&-\ft{1}{2} \ve^{ij} \bar{\e}_i\g_b \,{\tilde\F}^{ba}(V)
     \bpl 2\eta_{11} ( 2X\l_j +\phi\O_j^0 ) -i \eta_{1A}\O_j^A\bpr
     \nonumber\\
     &&+\ft12{i} \ve^{ij} \bar{\e}_i\g_b \,{\tilde\F}^{ba\, 0}
     \bpl 2X(2\eta_{11}\phi-g)\l_j +\eta_{11}\phi^2\O_j^0
     -i\eta_{1A}\phi\O_j^A -4i\der_I(Xb)\O_j^I\bpr  \nonumber\\
     &&+\ft12 {i} \ve^{ij} \bar{\e}_i\g_b \,{\tilde\F}^{ba\, A}
     \bpl i\eta_{1A} (2X \l_j +\phi\O_j^0) + 2\h_{AB} \O_j^B \bpr
     \nonumber\\
     && +i\ve^{ij}\bar{\e}_i\g_b\O_j^0 \, {\tilde{\hat 
B}}{}^{({\rm z})\, ba} 
     \nonumber\\
     && -\ft14 i \bar{\e}_i\g_b \, T^{ba\, ij} \Big[ 
     2 |X|^2 \bpl 2\eta_{11}\phi - {\rm Re}\, g \bpr\l_j \nonumber\\
     &&\hspace{24mm} +\bar{X}\bpl 2\eta_{11}\phi^2\O_j^0
     +\phi\bar{X}\der_{\bar{I}}\bar{g}\,\O_j^I
     -4i{\rm Re}[\der_I(Xb)]\O_j^I \bpr\Big]  \nonumber\\
     && +\ft{3}{2} i \bar\h^i\g^a \Big[ 2|X|^2 
     \bpl 2\eta_{11}\phi -  {\rm Re}\, g \bpr\l_i  \nonumber\\
     &&\hspace{12mm} +\bar{X}\bpl
     2\eta_{11}\phi^2\O_i^0 +\phi\bar{X}\der_{\bar{I}}\bar{g}\,\O_i^I
     -4i{\rm Re}[\der_I(Xb)]\O_i^I \bpr \Big] +{\rm h.c.}\,.
\err
The variation of the covariant fields
${\hat V}^{({\rm z})}_a$ and ${\hat B}^{({\rm z})}_{ab}$ equals
\brr 
\d {\hat V}^{({\rm z})}_a &=& i \ve^{ij} \bar{\e}_i \g_a
     \bpl 2 X \l_j + \phi \O^0_j \bpr^{({\rm z})}  
     + i \bar{\e}^iD_a\l_i -\ft18{i}\bar{\e}_i \g_a\s\cdot T^{ij}
     \l_j - \ft12{i} \bar{\h}^i \g_a \l_i + {\rm h.c.}\,, \nonumber\\
     \d {\hat B}^{({\rm z})}_{ab} &=& -4\bar{\e}^i\s_{ab}|X|^2 \bpl
     (2\eta_{11}\phi- {\rm Re}\, g )\l_i \bpr{}^{({\rm z})}  \nonumber\\
     && -2\bar{\e}^i\s_{ab}\, \phi^{({\rm z})} \bar{X} \bpl
     4\eta_{11}\phi \O_i^0 + \bar{X} \der_{\bar{I}}\bar{g}\,\O_i^I \bpr
     \\
     && - \ve^{ij}\bar{\e}_i\g_{[a} D_{b]}\bpl
     2X(2\eta_{11}\phi-g)\l_j  +\eta_{11}\phi^2\O_j^0
     -i\eta_{1A}\phi\O_j^A -4i\der_I(Xb)\O_j^I\bpr  \nonumber\\
     && + i \ve^{ij}\bar{\e}_i\g_{[a} {\hat V}^{({\rm z})}_{b]}
     \bpl 2\eta_{11}( 2X\l_j +\phi\O_j^0) -i\eta_{1A}\O_j^A \bpr 
     \nonumber\\
     && + i \bpl \h_{11} \F_{ab}(V) + \ft{1}{2} \h_{1A} \F_{ab}^A
     \bpr \bar{\e}^i\l_i \nonumber\\
     && - \ve^{ij} \bar\h_i \s_{ab} \bpl
     2X(2\eta_{11}\phi-g)\l_j +\eta_{11}\phi^2\O_j^0
     -i\eta_{1A}\phi\O_j^A -4i\der_I(Xb)\O_j^I\bpr + {\rm h.c.}\,
.\nonumber
\err
The same structure is repeated as one goes higher up in the 
central-charge
hierarchy. It was already observed in \cite{vt1} that the transformations
of the higher-$z$ fields involve objects both at the next and at the
preceding level. The transformations of the basic vector-tensor fields as
given in (\ref{momrules}) are special in this respect. They involve only
the next level as there is no lower level. The consistency of this is
ensured by the gauge transformations of the fields $V_\m$ and $B_{\m\n}$,
which allows for a truncation of the central charge hierarchy from below.

\setcounter{equation}{0}
\section{Invariant actions involving vector-tensor multiplets}
In this section we present the construction of invariant actions
for the vector-tensor
multiplet, using the multiplet calculus described in section 2.
We start by constructing a general linear multiplet depending on 
the vector-tensor 
fields and the background vector-multiplet components.
{}From this linear multiplet we construct the  associated supergravity
actions. Their dual description in terms of vector multiplets 
alone, which requires the use of field equations, is the
issue of the following section.

\subsection{The linear multiplet}
It is possible to form products of vector-tensor multiplets,
using the background vector multiplets judiciously, so as to form
$N=2$ linear multiplets.  
One starts by constructing the lowest component $L_{ij}$ of the 
linear multiplet in terms of vector-tensor fields as well as the 
background fields, which must 
have weights $w=2$ and $c=0$ and transform into a spinor doublet 
under $Q$-supersymmetry. We also note that 
$L_{ij}$ must transform as a real vector under chiral SU(2)
transformations.  The only vector-tensor component
which transforms under SU(2) is the fermion $\l_i$.
For the vector multiplets, only the fermions $\O_i^I$ and
the auxiliary fields $Y_{ij}^I$ transform nontrivially under SU(2).
Therefore, the most general possible linear multiplet must be 
based on an $L_{ij}$ of the following form
\brr 
  L_{ij} &=& X{\cal A}\,\bar{\l}_i\l_j
     +\bar{X}\bar{{\cal A}}\,\ve_{ik}\ve_{jl}\bar{\l}^k\l^l 
      +X{\cal B}_I\,\bar{\l}_{(i}\O_{j)}^I
     +\bar{X}\bar{{\cal B}}_{\bar{I}}\,
     \ve_{ik}\ve_{jl}\bar{\l}^{(k}\O^{Il)} \nonumber\\
     & & +{\cal C}_{IJ}\,\bar{\O}_i^I\O^J_j
     +\bar{{\cal C}}_{\bar{I}\bar{J}}\,
     \ve_{ik}\ve_{jl}\bar{\O}^{Ik}\O^{Jl} 
      +{\cal G}_IY_{ij}^I \,,
\label{ansatz}\err
where ${\cal A}$, ${\cal B}_I$, ${\cal C}_{IJ}$
and ${\cal G}_I$ are functions of $\phi$,  $X^I$ and
$\bar{X}^I$. In this section the index $I$ does not take the 
value $I=1$. In order that $L_{ij}$ has weights $w=2$ and $c=0$, 
the functions $\cal A$ and ${\cal G}_I$ must have weights 
$w=c=0$, while ${\cal B}_I$ and ${\cal C}_{IJ}$ have weights 
$w=-c=-1$.
Obviously, the reality condition on $L_{ij}$ requires that
${\cal G}_I$ be real.  As before, we suppress the superscript 
zeroes of the central-charge vector multiplet for the sake of clarity.
We also expect the linear multiplet to transform only under the central
charge and not under the gauge transformations associated with the
other vector multiplets, but this is not important for most of 
the construction.  

Requiring that $L_{ij}$ transforms into a spinor doublet
as indicated in (\ref{linear}), puts stringent requirements on 
each of the functions 
${\cal A}(\phi,X^I,\bar{X}^I)$,
${\cal B}_I(\phi,X^I,\bar{X}^I)$,
${\cal C}_{IJ}(\phi,X^I,\bar{X}^I)$
and ${\cal G}_I(\phi,X^I,\bar{X}^I)$,
which take the form of coupled first-order, linear differential
equations. These equations are exactly the same as in the rigid 
case, which were given in
\cite{vt2}. We will not repeat them here but immediately present 
their solution, which is a 
linear combination of three distinct solutions, each with
an independent physical interpretation. The most interesting of these
is given as follows,
\brr [{\cal A}{}]_1 &=&
     \eta_{11}(\phi+i\zeta)
     -\ft12 g \,, \nonumber\\
     {}[{\cal B}_I{}{}]_1 &=&
     -\ft12(\phi+i\zeta)\der_I g
     -2i\der_I b \,,\nonumber\\
     {}[{\cal C}_{IJ}]_1 &=&
     -\ft12 i(\phi+i\zeta)\der_I\der_J(Xb) \,, \nonumber\\
     {}[{\cal G}_I{}]_1 &=& {\rm Re}\bl
     [\ft13\eta_{11}(\phi+i\zeta)^3
     -\ft12 i\zeta(\phi+i\zeta)g]\d_I{}^0
     +\ft12(\phi+i\zeta)X\der_I(g\phi+4ib)\br \,,
\label{first}\err
where
\bq \zeta(\phi, X^I,\bar X^I)=\frac{{\rm Im}(\phi g+4ib)}
    {2\eta_{11}\phi-{\rm Re}\,g} \,.
\label{zetadef}\eq
In terms of the action, which will be discussed shortly,
this solution provides the couplings which involve the
vector-tensor fields.  The remaining two solutions,
which we discuss presently, give rise either to a total divergence
or to interactions which involve only the background fields.
The latter of these correspond to previously known results.
The second solution takes the form,
\brr {}[{\cal A}{}]_2 &=&
     i\eta_{11}\zeta'
     -i\a \,, \nonumber\\
     {}[{\cal B}_I{}]_2 &=&
     -\ft12 i\zeta'\der_I g
     -2i\der_I\g \,,
     \nonumber\\
     {}[{\cal C}_{IJ}{}]_2 &=&
     \ft12\zeta'\der_I\der_J(Xb) \,,\nonumber\\
     {}[{\cal G}_I{}]_2 &=& {\rm Re}\bl
     2i X \phi \der_I \g + \ft i2 \z' X \phi \der_I g - 2 \z' \der_I
     (Xb)\br\,,
\label{second}\err
where $\g=\ft14 i\a_A{X^A}/{X}$ is a holomorphic homogeneous 
function of the background scalars $X^A$ and $X^0$;  $\a$ and 
$\a_A$ are arbitrary real parameters. Furthermore  
\bq \zeta'(\phi, X^I,\bar X^I)=\frac{2\a\phi+4{\rm Re}\,\g}
    {2\eta_{11}\phi-{\rm Re}\,g} \,.
\label{zpdef}
\eq
Note that this solution could be concisely included into the first
solution by redefining $g\to g+2i\a$ and $b\to b+\g$.
In fact, this second solution indicates that
the functions $g$ and $b$ are actually
defined modulo these shifts.
In terms of the action, this ambiguity is analogous
to the shift of the theta angle in an ordinary Yang-Mills theory.

The third and final solution is given by
\brr {}[{\cal A}{}]_3 &=& 0 \,,
     \nonumber\\
     {}[{\cal B}_I{}]_3 &=& 0 \,,
     \nonumber\\
     {}[{\cal C}_{IJ}{}]_3 &=&
     -\ft18 i\der_I\der_J(f(X)/X) \,,
     \nonumber\\
     {}[{\cal G}_I{}]_3 &=&
     -\ft12{\rm Im}\,\der_I(f(X)/X ) \,.
\label{third}\err
Where $f(X)$ is a holomorphic function of $X^0$ and $X^A$,
of degree 2. In terms of the action, this solution
corresponds to interactions amongst the background vector
multiplets alone.  Since the possible vector multiplet
self-couplings have been fully classified, this solution
does not provide us with new information.  The function $f(X)$
provides the well-known holomorphic prepotential for describing
the background self-interactions.

All solutions have in common that they are homogeneous functions 
of $X^I$ and $\bar X^I$: $\cal A$ and ${\cal G}_I$ are of degree 0 
and ${\cal B}_I$ and ${\cal C}_{IJ}$ are of degree $-1$. This is a 
result of the fact that the field $\phi$ has $w=0$. Furthermore we 
note the identities,
\bq
X^I\,{\cal B}_I = X^I\,{\cal C}_{IJ}=0\,,
\eq
which ensure that $L_{ij}$ is invariant under $S$-supersymmetry, 
in accord with (\ref{linear}). 

Now that we have determined the scalar triplet $L_{ij}$,
in terms of the specific functions
${\cal A}(\phi,X^I,\bar{X}^I)$,
${\cal B}_I(\phi,X^I,\bar{X}^I)$,
${\cal C}_{IJ}(\phi,X^I,\bar{X}^I)$, and
${\cal G}_I(\phi,X^I,\bar{X}^I)$
given above, we can generate
the remaining components of the linear multiplet,
$\varphi_i,\, G$, and $E_\mu$ by varying (\ref{ansatz})
with respect to supersymmetry. Given the complexity of the
transformation rule for $\l_i$ found in (\ref{momrules}),
it is clear that a fair amount of work is involved in
carrying out this process. However, since we are only interested in the
bosonic part of the action,
we are only interested in the bosonic part of $E_a$ and $G$, viz.
(\ref{linaction}).

The higher components of the linear multiplet are then given by
\brr \varphi^i &=& -\bar{X}(\Dslash\phi+i \hat V\!\!\llap/\,\,{}^{({\rm z})})
     (\bar {\cal A}\l^i+\ft12 \bar{\cal B}_{\bar{I}}\O^{Ii})
     +{\cal G}_I\Dslash\O^{Ii} \nonumber\\
     & & -\ft{i}{2}\ve^{ij}\s\cdot
     (\F(V)-i\phi\F^0)({\cal A}\l_j+\ft12 {\cal B}_I\O^I_j) \nonumber\\
     & & +\ft12\ve^{ij}\s\cdot\F^I(X{\cal B}_I\l_j+2{\cal C}_{IJ}\O_j^J)
     \nonumber\\
     & & -\Dslash\bar{X}^I(\bar{X} \bar{\cal B}_{\bar{I}}\l^i
     +2\bar{\cal C}_{\bar{I}\bar{J}}\O^{Ji}) \nonumber\\
     & & -|X|^2\phi^{({\rm z})}\ve^{ij}(2{\cal A}\l_j+{\cal B}_I\O_j^I)
     \nonumber\\
     & & +\ft12 Y^{Iij}\Big(
     (\der_\phi {\cal G}_I)\l_j+(\der_J {\cal G}_I)\O_j^J\Big)
     +{\rm 3\,fermion\ terms}\,, \nonumber\\
     G &=&\bar X \bar{\cal A}\,(D_a\phi+i\hat V^{({\rm z})}_a)
     (D^a\phi+i{\hat V}^{a({\rm z})}) \nonumber\\
     & & +2\bar{X}\bar{\cal B}_{\bar{I}}\,D_a\bar{X}^I
     (D^a\phi+i{\hat V}^{a ({\rm z})}) \nonumber\\
     & & +4\bar{\cal C}_{\bar{I}\bar{J}}\,D_a\bar{X}^I\,D^a\bar{X}^J
     -2{\cal G}_I \,D_aD^a\bar{X}^I \nonumber\\
     & & +\frac{1}{4X}(\F(V)^--i\phi\F^{0-})_{ab}
     \big({\cal A}(\F(V)^--i\phi\F^{0-})
     +2i X{\cal B}_I\F^{I-}\big)^{ab} \nonumber\\
     & & -{\cal C}_{IJ}\F^{I-}_{ab}\F^{J-ab}
     -4\bar{X}|X|^2 {\cal A} (\phi^{({\rm z})})^2 \nonumber\\
     & & -\ft14(\der_{(I}{\cal G}_{J)}+X^{-1}P_{(I}\,\der_\phi 
    {\cal G}_{J)})\, 
     Y_{ij}^IY^{Jij}\nonumber\\
     & & -\ft12 {\cal G}_I \,\F^{I+}_{ab}\, T^{ab}_{ij}\ve^{ij} 
     +{\rm fermion\ terms}\,,\nonumber\\
     E_a &=& {\rm Re}\bpl-4|X|^2\phi^{({\rm z})}
     (\bar{\cal A}\,(D_a\phi+i {\hat V}_a^{({\rm z})}) + {\cal 
    B}_I \,D_a X^I)
     \nonumber\\
     & & \hspace{.3in} -2i(D^b\phi + i {\hat V}^{({\rm z})\, b})({\cal
     A}\,(\F(V)^-_{ab} -i \phi\F_{ab}^{0-})
     +iX {\cal B}_I\F_{ab}^{I-}) \nonumber\\
     & & \hspace{.3in} -2D^b X^I(i{\cal B}_I \,(\F(V)_{ab}^-
     -i \phi\F_{ab}^{0-})
     -4{\cal C}_{IJ}\,\F_{ab}^{J-}) \nonumber\\
     & & \hspace{.3in} -2 {\cal G}_I \,D^b(\F_{ab}^{-I}
     -\ft14  \bar{X}^I T^{ij}_{ab} \ve_{ij}) \bpr + {\rm 
    fermion\ terms} \,. 
\label{lincomponents}
\err
Here we used the notation
\bq 
  P_I = -\ft12\phi\,\d_I{}^0 +i {{\rm Im}\Big(
     \phi X\der_I g +4i\der_I(Xb)\Big)\over  2(2\eta_{11} \phi - 
   {\rm Re}\, g)}  \,.
\label{epdef}
\eq
The appearance of terms containing $T_{ab}^{ij}$ may seem strange because
this field does not appear in the transformation rules for $\l_i$ and
$\O_i$. However, this field appears in the variation of $\Dslash \O_i$
and in the Bianchi identities for $\F^I_{ab}$, which have to be 
used to obtain $G$ and $E_a$. Having derived the complete linear 
multiplet we can construct the action.

\subsection{The action}
Now we want to use the linear multiplet components derived above
in the action formula (\ref{linaction}). Since this linear multiplet
transforms under the central
charge we need to use the central-charge vector multiplet in the action
formula, as explained in section 2. This yields an action that is both
invariant under local supersymmetry and local gauge transformations.
Carrying out this calculation we note the following term in 
Langrange density, 
\bq 
{\cal L} = 4 e \bar X{{\cal C}}_{IJ}\, D^a X^I D_a X^J
     -2e {\cal G}_I\, X\,D_a D^a \bar{X}^I \cdots  \,,
\eq
which we rewrite by splitting off a total derivative. This 
leads to derivatives of the function ${\cal G}_I$, which we 
rewrite using its explicit form (or the differential equations of 
which it is a solution).  After this manipulation, the bosonic 
terms of the full action read,
\brr 
   e^{-1} {\cal L} &=& -2 {\cal G}_I\, X \bar{X}^I (\ft16 {\cal 
R} - D)\nonumber\\ 
     & & + |X|^2{\cal A}\,(\der_\m \phi -i{\hat V}^{({\rm z})}_\mu)^2
     +2|X|^2{\cal B}_I\,D^\mu X^I
     (\der_\mu\phi -i{\hat V}_\mu^{({\rm z})}) \nonumber\\
     & & - 4  X {{\cal C}}_{IJ}\,
     D^\mu X^I D_\mu \bar{X}^J
     -2 \bar{X}(X {\cal B}_I + {\cal A}\, P_I) \der_\mu \phi \,
D^\mu X^I\nonumber\\ 
     & &-2 X ({\cal B}_I \,P_J D_\mu X^I  +
     \bar {\cal B}_{\bar I}\, \bar P_{\bar J}\,D_\mu \bar{X}^I)\, D^\mu
     \bar{X}^J + 2 {\cal G}_I\, D_\mu X \,D^\mu \bar{X}^I \nonumber\\
     & & + {\cal A}\,(\F(V)^{-\, \mu\nu}-i\phi\F^{-\, 0\mu\nu})\bpl
     \ft14(\F(V)_{\mu\nu}^--i\phi\F^{-\, 0}_{\mu\nu})
     +iW^{0}_{\mu}(\der_\nu \phi -i{\hat V}_\nu^{({\rm z})})\bpr \nonumber\\
     & & +iX{\cal B}_I\,\F^{-\, I \mu\nu}\bpl
     \ft12(\F(V)_{\mu\nu}^--i\phi\F^{-\, 0}_{\mu\nu})
     +iW^0_\mu (\der_\nu \phi - i{\hat V}_\nu^{({\rm z})})\bpr \nonumber\\
     & & +i{\cal B}_I\,(\F(V)^{-\, \mu\nu} - i\phi \F^{-\, \mu\nu})W^0_\mu
     D_\nu X^I\nonumber\\
     & & -{\cal C}_{IJ}\,\F^{I-\mu\nu}\bpl X \F^{J-}_{\mu\nu} + 4 W^0_\mu
     D_\nu X^J\bpr
     \nonumber \\
     & & -|X|^2 {\cal A}\,(W_\mu^{0}\,W^{\m\,0} + 
   4|X|^2)(\phi^{({\rm z})})^2 \nonumber\\ 
     & & -\ft14(X\,\der_{(I}{\cal G}_{J)}
     +P_{(I}\,\der_\phi{\cal G}_{J)})Y_{ij}^IY^{Jij}
     -\ft14{\cal G}_I\,Y_{ij}^0 Y^{Iij}\nonumber\\
     & &- \ft12 {\cal G}_I X \F^{I+}_{ab}\, T^{ab}_{ij}\ve^{ij}  
    +{\cal G}_I\,W^{0}_a \,D_b( \F^{-I\,ab} 
    - \ft14 \bar{X}^I T^{ab\,ij} \ve_{ij} ) + {\rm h.c.}\,,
\label{lagrangian}
\err
where we have made the terms proportional to $W^0_\m$ in the 
covariant derivatives explicit. 
The above result  describes the coupling of a vector-tensor 
multiplet to 
$n$ vector multiplets. Note that each term involves a factor
of the functions ${\cal A}(\phi,\,X^I,\,\bar{X}^I)$,
${\cal B}_I(\phi,\,X^I,\,\bar{X}^I)$,
${\cal C}_{IJ}(\phi,\,X^I,\,\bar{X}^I)$
${\cal G}_I(\phi,\,X^I,\,\bar{X}^I)$
or $P_I(\phi,\,X^I,\,\bar{X}^I)$, which were given explicitly
in the previous section.

This form of the action would be a suitable starting point to consider the
breaking of superconformal gravity into Poincar\'e gravity. An additional
compensator e.g. a hypermultiplet would be needed to be able to define a
gauge for the dilatations. The procedure
would then be completely analogous to the case described in \cite{DWLVP}.
However, it is not the purpose of this paper to go into the details of this.

In the general case described above, the functions
${\cal A}$,
${\cal B}_I$,
${\cal C}_{IJ}$ and
${\cal G}_I$,
which define the Lagrangian are linear superpositions
of three distinct terms, one of which describes the
local couplings of the vector-tensor multiplet components,
another which is a total derivative,
and one which codifies the self-interactions of the background.
As a result of this, the Lagrangian (\ref{lagrangian}) can
be written as a sum of three analogous pieces: a vector-tensor
piece, a total-derivative piece, and a background piece.

Now that we have given the action in terms of the functions
${\cal A}$,
${\cal B}_I$,
${\cal C}_{IJ}$ and
${\cal G}_I$, it is instructive to give the
solutions for the two inequivalent representations described in section
\ref{s:vttrans}.
\vspace{.1in}

\noindent{\it The nonlinear vector-tensor multiplet:}\\
As described above, when the parameter $\eta_{11}$ does not vanish,
the tensor field involves a coupling to the Chern-Simons form
$V\wedge {\rm d}V$, which is quadratic in terms of vector-tensor fields.
Consequently, the corresponding transformation rules contain
significant nonlinearities.  As was shown in \cite{vt2}, in this case
it is possible to remove the parameter $\eta_{1A}$, and therefore
the $V\wedge {\rm d} W^A$ Chern-Simons couplings.
Without loss of generality, we then define
$\eta_{11}=1$ and $\eta_{1A}=0$. In this case the functions
${\cal A}(\phi,X^I,\bar{X}^I)$,
${\cal B}_I(\phi,X^I,\bar{X}^I)$,
${\cal C}_{IJ}(\phi,X^I,\bar{X}^I)$, and
${\cal G}_I(\phi,X^I,\bar{X}^I)$
which define the linear multiplet and, more importantly,
the vector-tensor Lagrangian (\ref{lagrangian}) are given by
the following expressions
\brr {\cal A} &=& \phi+i\phi^{-1}(b+\bar{b})\,, \nonumber\\
     {\cal B}_I &=& -2i\der_Ib \,,\nonumber\\
     {\cal C}_{IJ} &=& -\ft12 i(\phi
     +i\phi^{-1} (b+\bar{b})\der_I\der_J(Xb)
     -\ft18 i\der_I\der_J(X^{-1}f) \,,\nonumber\\
     {\cal G}_I &=& {\rm Re}\bpl\ft13\phi^3\,\d_I{}^0
     +2i\phi X\der_I b
     -2\phi^{-1}(b+\bar{b})\,\der_I(Xb)\bpr
     -\ft12{\rm Im}\,\der_I\bpl X^{-1}f\bpr \,.
\label{nonlin}\err
For the sake of clarity, we have absorbed the parameters
$\a$ and $\a_A$ into the functions $b$ and $g$ in the manner
described immediately after equation (\ref{zpdef}).
Substituting these functions in the Lagrangian (\ref{lagrangian}),
it is easy to see that the action contains, besides the total derivative and
terms that depend only on the background vector multiplet fields,
a cubic part and a linear part in vector-tensor fields. This is 
the immediate generalization to a background with more than one 
vector multiplet of the Lagrangian described in \cite{vt1}.
\vspace{.1in}

\noindent{\it The linear vector--tensor multiplet:}\\
As described previously, if $\eta_{11}=0$, implying the absense of the
$V\wedge {\rm d}V$ Chern-Simons coupling,
we obtain a vector-tensor multiplet which is
distinct from the nonlinear case just discussed.  In this case, it
is not possible to perform a field redefinition to remove all
of the $\eta_{1A}$ parameters and  the supersymmetry
transformation rules are linear in terms of the
vector-tensor component fields.
The functions  ${\cal A}(\phi,X^I,\bar{X}^I)$,
${\cal B}_I(\phi,X^I,\bar{X}^I)$,
${\cal C}_{IJ}(\phi,X^I,\bar{X}^I)$, and
${\cal G}_I(\phi,X^I,\bar{X}^I)$
which define the linear multiplet and, more importantly,
the vector-tensor Lagrangian (\ref{lagrangian}) are now given by
the following expressions
\brr {\cal A} &=& -\ft12 g \,,\nonumber\\
     {\cal B}_I &=& -\frac{1}{g+\bar{g}}\bpl
     \phi\bar{g}\der_Ig
     +2i(g+\bar{g})\derbar_I(b+\bar{b})\bpr\,, \nonumber\\
     \nonumber\\
     {\cal C}_{IJ} &=&
     -\frac{1}{g+\bar{g}}
     \bpl i\phi\bar{g}+2(b+\bar{b})\bpr\der_I\der_J(Xb)
     -\ft18 i\der_I\der_J(X^{-1}f)\,, \nonumber\\
     {\cal G}_I &=& \frac{1}{g+\bar{g}}{\rm Re}\bl
     \phi\bar{g}X\der_I(\phi g+4ib)
     -2i(b+\bar{b})\der_I[X(\phi g+4ib)]\br \,.
\err
As above, for the sake of clarity we have absorbed the parameters
$\a$ and $\a_A$ into the functions $b$ and $g$ in the manner
described immediately after equation (\ref{zpdef}).
Substituting these functions into the Lagrangian (\ref{lagrangian}), one
obtains a Lagrangian that contains, besides the total derivative terms and
a part that depends exclusively on the background mentioned 
above, a quadratic part and a linear part in vector-tensor fields.

\setcounter{equation}{0}
\section{Dual versions of vector-tensor actions}
As we already mentioned in the introduction, 
a vector-tensor multiplet is classically equivalent to a vector
multiplet. The theory which we have presented, involving
one vector-tensor multiplet and $n$ vector multiplets
is classically equivalent to a theory involving $n+1$ vector
multiplets. Since these latter theories are well understood, it
is of interest to determine what subset of vector multiplet
theories are classically equivalent to vector-tensor theories.
Furthermore, low-energy effective string Lagrangians with $N=2$
supersymmetry are usually described in terms of vector multiplets, such that
by going to the vector multiplet language one can more easily verify 
which string theories are described by the vector-tensor multiplets
we constructed above. 
A significant restriction along these lines has to do with the
K\"ahler spaces on which the scalar fields of the theory may live.
In the case of $N=2$ vector multiplets these consist of ``special 
K\"ahler" spaces, and the associated geometry is 
known as special geometry. For the case of effective
Lagrangians corresponding to heterotic $N=2$ supersymmetric
string compactifications, this space must contain, at least at 
weak string coupling, an SU(1,1)/U(1) coset factor parametrized in terms of 
the complex scalar corresponding to the axion/dilaton complex. 
According to a well-known theorem \cite{FVP} this uniquely 
specifies the special K\"ahler space. 
 
Perhaps not too surprisingly, the observations made in \cite{vt2} are 
not altered by going to local supersymmetry. Thus we will find that 
the vector-tensor multiplets we have been studying in the present article, 
fail to exhibit the SU(1,1)/U(1) factor, at least if one insists 
that it is the vector-tensor  scalar and tensor field (the latter after 
a duality transformation, to be discussed below) that parametrize this 
subspace. Therefore it is impossible to associate this scalar 
and the tensor field with the (perturbative) heterotic   
dilaton-axion complex. However, they do play a natural role in 
the description of the non-perturbative heterotic string effects 
we alluded to in the introduction. 

One goes about constructing the dual vector multiplet formulation,
in the usual manner, by introducing a Lagrange multiplier
field $a$, which, upon integration, enforces the Bianchi identity
on the field strength $H_\mu$.
The relevant term to add to the Lagrangian is therefore
\brr e^{-1} {\cal L}(a) &=&
     a\,   D_\m H^\m  \nonumber\\
     &&+  \ft14 i a \Big[\eta_{11}\, \tilde{\cal F}_{\m\n}(V)\,
     {\cal F}^{\m\n}(V) +\eta_{1A}\,\tilde {\cal F}_{\m\n}(V)\,
     {\cal F}^{\m\n A} +\eta_{AB}\, \tilde{\cal F}_{\m\n}^{A}\,
     {\cal F}^{\m\n B} + 2 \tilde \F^{0}_{\m\n}\hat B^{\m\n\,({\rm 
     z})}\Big]   \nonumber\\    
     &&+\ft1{16}i a\Big[T_{ij}^{\m\n}\Big(
     2\eta_{11}\, \phi X\,{\cal F}_{\m\n}(V) +\eta_{1A}(X^A \,
     {\cal F}_{\m\n}(V) +i \phi X {\cal F}_{\m\n}^A) +2\eta_{AB}\, 
     X^A\,{\cal F}_{\m\n}^B \nonumber \\
     && \hspace{1.8cm}   + 2X \hat B^{({\rm z})}_{\m\n}+X  \, 
     \F^{0}_{\m\n} (\eta_{11} 
     \phi^2 -\phi\,g -4ib)\Big)  - {\rm h.c.}\Big] \,.
\label{aterm}\err
Note that we dropped the explicit fermionic terms, as we will do 
in the remainder of this section. Including the Lagrange 
multiplier term, we treat $H_\mu$ as unconstrained and integrate
it out in the action, thereby trading the single on-shell degree
of freedom represented by $B_{\mu\nu}$ for the real scalar $a$.
Doing this, we obtain a dual theory involving only
vector multiplets.  To perform these operations, it is instructive to
note that all occurences of $H_\mu$ in
(\ref{lagrangian}) and (\ref{aterm}) are most conveniently written
in terms of $\hat V_\mu^{({\rm z})}$, which can be done using 
(\ref{vzbz}). Because we are suppressing the fermions in what 
follows, we will henceforth drop the caret on $V_\m^{(\rm z)}$. 
All such terms can then be collected, and written as follows,
\bq 
  {\cal L}( V_\mu^{({\rm z})})= \ft14e(2\eta_{11}\phi-{\rm Re}\,g) \bpl
    W^{0\,\mu} W^{0\,\nu}-(W_\l^0 W^{0\,\l}+4|X|^2)g^{\mu\nu}\bpr
    \bpl V_\mu^{({\rm z})} V_\nu^{({\rm z})}
   -2V_\mu^{({\rm z})}\der_\nu(a-\zeta)\bpr   \,,
\label{lvz}\eq
where $\zeta$ was defined in (\ref{zetadef}).
It is interesting how the terms involving $V_\mu^{({\rm 
z})}$ factorize into the form given in (\ref{lvz}). The equation 
of motion for $H_\mu$ is conveniently written in terms of 
$V_\mu^{({\rm z})}$, which follows immediately from (\ref{lvz}). 
It is given by the following simple expression,
\bq V_\mu^{({\rm z})}=\der_\mu(a-\zeta) \,.
\label{vzeom}
\eq
We also impose the equations of motion for the auxiliary fields, 
$\phi^{({\rm  z})}=Y_{ij}^I=0$ (up to fermionic terms). After substituting  
these solutions, we manipulate the 
result into the familiar form for the bosonic Lagrangian
involving vector multiplets,
\brr e^{-1} {\cal L} &=& \ft12i (F_I \bar X^I - X^I \bar F_I) \bpl 
-\ft16 {\cal R} + D  \bpr 
     +\ft12 i\big({\cal D}_\mu F_I\,{\cal D}^\mu\bar{X}^I
     -{\cal D}_\mu X^I\,{\cal D}^\mu\bar{F}_I\big)\nonumber \\
     &&- \ft18 i {\bar F}_{IJ} F_{\mu\nu}^{+ I} F^{+\mu\nu\, J} 
      - \ft1{16}i(F_I - X^J\bar F_{JI}) F_{\mu\nu}^{+ I} 
     T^{\mu\nu}_{ij} \ve^{ij}    \nonumber\\
     &&+ \ft{1}{128}i(F_I - X^J\bar F_{JI})X^I \bpl T_{\mu\nu ij} 
     \ve^{ij} \bpr^2 +  {\rm h.c.}  \, ,
\label{aform}
\err
characterized by a holomorphic function $F(X^0,X^1,X^A)$, which 
is homogeneous of degree two. Here the field strengths are 
equal to $F_{\m\n}^I = 2\der_{[\m}W^I_{\n]}- g 
f_{JK}{}^{\!I}W_\m^JW_\n^K$. 
In (\ref{aform}), a subscript $I$ denotes differentiation
with respect to $X^I$. The natural bosons in the dual theory are 
found to be 
\brr 
   X^1 &=& X^0\Big((a-\zeta)+i\phi\Big)\,, \nonumber\\
     W_\mu^1 &=& V_\mu+(a-\zeta)W_\mu^0 \,,
\label{defW1}
\err
and one can check that these transform as components
of a common vector multiplet.
For the general case, the dual theory obtained
in this manner is described by the following
holomorphic prepotential,
\brr 
   F(X^0, X^1,X^A) &=& -\frac{1}{X^0}\bpl
    \ft13\eta_{11}X^1X^1X^1
    +\ft12\eta_{1A}X^1X^1X^A
    +\eta_{AB}X^1X^AX^B \bpr \nonumber\\
    & & -\a X^1X^1
    +\a_A X^1X^A
    +f(X^0,X^A) \,.
\label{prepot}\err
The quadratic terms proportional to $\a$ and $\a_A$ (defined in 
section~4.1) give rise to total derivatives since
their coefficients are real.  The term involving
the function $f(X^0,X^A)$ represents the self-interactions
of the background vector multiplets.
The first three terms in (\ref{prepot})
encode the couplings of the erstwhile vector-tensor fields,
$\phi$ and $a$, and it is these which we are most interested in.
As mentioned above, it is relevant to investigate whether the
K\"ahler space described by this prepotential function can 
contain an SU(1,1)/U(1) factor parametrized by the field 
$X^1/X^0$. According to the theorem of \cite{FVP}, this requires 
that $X^1/X^0$ appears linearly in the prepotential. This is 
obviously not the case for (\ref{prepot}), as we have quadratic 
and cubic terms which cannot be removed by absorbing some of the 
other fields into the would-be dilaton field $X^1/X^0$.  As 
discussed earlier in this paper, the 
best one can do is to remove {\it either} $\eta_{11}$ or
$\eta_{1A}$.  There exists an obstruction to removing both
of these.  We recall that these parameters are related to the
Chern-Simons couplings of the tensor field in the dual
formulation.  The obstruction to removing the unwanted terms
in the prepotential derives from the inability to formulate
an interacting off-shell vector-tensor theory without any such Chern-Simons
couplings.

In the present supergravity context it is important to note that the 
duality transformation we just described, does not interfere with the 
fields of the Weyl multiplet. This can be seen by 
nothing that (\ref{lvz}), (\ref{vzeom}) and (\ref{defW1}) are 
completely identical to the relations found in \cite{vt2} in the 
rigid supersymmetric case. This implies that the Weyl multiplet is 
not involved in the duality transformation and can be kept 
off-shell. The vector multiplets 
are not realized off-shell after the duality transformation, but 
the auxiliary fields $Y^I_{ij}$ can be reinstated 
afterwards. In this respect it is instructive to 
compare our results to the  
analysis performed in \cite{siebel}. Here the most general 
vector-multiplet theories admitting a (reverse) 
dualization into an antisymmetric tensor theory, were considered. 
They were found to precisely comprise the 
cases described here, plus the $\eta_{11}=0$, $\eta_{1A}=0$ case which is 
relevant for weakly coupled heterotic strings. However, in this last 
case the dualization into an antisymmetric tensor theory can no 
longer be carried out with the Weyl multiplet as a 
spectator. In particular, one is forced to first eliminate the U(1) 
chiral gauge field $A_\m$, which in the Poincar\'e theory plays 
the role of an auxiliary field.

Irrespective of these considerations, we note that the results we obtained in
this article are
a concise description of two very different situations.  As described
in detail in section 3, depending on whether the parameter
$\eta_{11}$ is vanishing or not, indicating the absence or presence,
respectively, of a $V\wedge {\rm d}V$ Chern-Simons coupling to the tensor
field, the theory takes on very distinct characters.  It is instructive
then, to summarize our results independently for each of these two
cases.

For the nonlinear vector-tensor multiplet, we
obtain a dual description involving only vector multiplets,
characterized by the following holomorphic
prepotential,
\brr F &=& -\frac{X^1}{X^0}\bpl
    \ft13 \eta_{11} X^1X^1 +\eta_{AB}X^AX^B \bpr
    -\a X^1X^1
    +\a_A X^1X^A
    +f(X^0,X^A) \,.
\err
As already mentioned above, the quadratic terms
proportional to $\a$ and $\a_A$ represent
total derivatives, and the last term involves the background
self-interactions.  Notice that in this case the
prepotential is cubic in $X^1$. No higher-dimensional tensor theory
is known that gives rise to this coupling.

For the linear vector-tensor multiplet the dual description 
in terms of only vector multiplets is characterized by the following
prepotential,
\brr F &=& -\frac{X^1}{X^0}\bpl
    \ft12\eta_{1A}X^1X^A +\eta_{AB}X^AX^B \bpr
    -\a X^1X^1
    +\a_A X^1X^A
    +f(X^0,X^A)\,. \nonumber\\
\err
Again, as discussed above, the quadratic terms
involving $\a$ and $\a_A$ represent
total derivatives, while the last term involves the background
self-interactions.
Notice that in this case the prepotential has a term quadratic in
$X^1$, which cannot be suppressed. Such a term also arises from
the reduction of six-dimensional tensor multiplets to four dimensions.
In that case, the presence of the quadratic term is inevitable, because it
originates from the kinetic term of the tensor field 
\cite{FerMinSag}. Observe that we have at least
three abelian vector fields coupling to the vector-tensor
multiplet, namely $W^0_\m$, $W^1_\m$ and $\eta_{1A}W^A_\m$.

The work presented in this paper represents an exhaustive
analysis of the $N=2$ vector-tensor multiplet coupled to 
supergravity and a number of background vector
multiplets. One of these vector multiplets provides the gauge 
field that couples to the central charge. Although we considered 
only a single vector-tensor multiplet, our methods can be 
straightforwardly applied to theories where several of these 
multiplets are present. 
We have presented the complete
and general superconformal transformation rules in this context, and
have shown that these actually include two distinct cases,
one of which is nonlinear in the vector-tensor components,
and the other of which is linear. The difference between these two 
cases is encoded in the the coefficients of the Chern-Simons 
couplings, denoted by $\eta_{IJ}$. 
Furthermore we have constructed a supersymmetric action for
this system, and exhibited its bosonic part. The
dual descriptions in terms of vector multiplets
have been obtained, and the respective prepotential
functions exhibited.  

\vspace{1cm}

\noindent
{\bf Acknowledgement}

\noindent
We thank F. Brandt, N. Dragon, E. Ivanov, S.M. Kuzenko, B.A. 
Ovrut and E. Sokatchev for informative discussions.\\
Work supported by the European Commission TMR programme  
ERBFMRX-CT96-0045, in which P.~C. is    
associated to Leuven, B.~d.W. and B.~K. to Utrecht, M.~F. to 
Berlin and P.~T. to Torino. \\
P.~C. and R.~S. thank the FWO (Belgium) and 
B.~K. and M.~F. the FOM (The Netherlands) for financial support.
P.~T. is a postdoctoral fellow in the above TMR programme.  

\pagebreak

\appendix
\setcounter{equation}{0}
\begin{section}{Conventions and definitions}

Throughout the article we use $\mu ,\nu , \cdots = 0,1,2,3$ to denote
curved indices, and $a,b, \cdots = 0,1,2,3$ for local Lorentz indices. 
Our (anti)symmetrizations are always with weight one, so e.g. 
\bq
[ab] = \ft12 ( ab-ba )\,,
\eq
We take
\bq
\g_a \g_b = \h_{ab} + 2 \s_{ab}\,, \qquad 
\g_5 = i \g_0\g_1\g_2\g_3\,,
\eq
where $\h_{ab}$ is of signature $(-+++)$. The complete antisymmetric
tensor satisfies
\bq
\ve^{abcd} = e^{-1} \ve^{\mu\nu\l\s} e_\mu^a e_\nu^b e_\l^c 
e_\s^d\,,   \qquad
\ve^{0123} = i \,,
\eq
which implies 
\bq \s_{ab} = -\ft12\ve_{abcd}\s^{cd}\g_5\,.
\eq
The dual of an antisymmetric tensor field $F_{ab}$ is given by
\bq
\tilde{F}_{ab} = \ft12\ve_{abcd}F^{cd}       \,.
\eq
and the (anti)selfdual part of $F_{ab}$ reads
\bq
 F^{\pm}_{ab} = \ft12(F_{ab}\pm\tilde{F}_{ab})  \,.
\eq
Note that under hermitian conjugation ($h.c.$) selfdual becomes antiselfdual
and vice-versa. Any SU(2) index $i$ or any quaternionic
index $\a$ changes position under ${\rm h.c.}$, for instance
\bq (T_{ab\, ij})^* = T_{ab}^{ij} \,,
\qquad \qquad (A_i^\a)^* = A^i_\a\,.
\eq
The superconformal algebra consists of
general coordinate, local Lorentz,
dilatation, special conformal,
chiral U(1) and SU(2), and $Q$- and $S$-supersymmetry transformations. When
vector and/or vector-tensor multiplets are present additional gauge
symmetries must be included. A covariant general coordinate transformation
is defined as follows
\bq \d^{(cov)}(\xi)=\d_{{\rm gct}}(\xi)
    +\sum_{A}\d_A (-\xi^\mu h_\mu(A)) \,,
\eq
where the sum is over all superconformal (except the g.c.t) and additional
gauge transformations,
each with parameter $-\xi^\mu h_\mu (A)$,
where $h_\mu(A)$ is the gauge field associated with $\d_A$.
The superconformal gauge fields are normalized as in \cite{DWLVP}.
\bq
\begin{array}{rclrcl}
h_\mu^{ab}(M) &=& \o_\mu^{ab}\,, \hspace{2cm}&h_\mu (D) &=& b_\mu 
\,,\\
h_\mu(U(1)) &=& A_\mu  \,, 
&{h_\mu}^i_{\,\, j}(SU(2)) &=& - \ft12 {{\cal V}_\mu}^i_{\,\, 
j}\,,\\
h_\mu^i(Q) &=& \ft12 \psi_\mu^i\,, & h_\mu^i(S) &=& \ft12 
\phi_\mu^i\,, \\
h_\mu^a (K) &=& f_\mu^a \,,
\end{array}
\eq
and
\bq
h_\mu ({\rm gauge}) = W_\mu^I, V_\mu, B_{\mu\nu}\,.
\eq
The symbol $D_\mu$ denotes a fully covariant derivative
and is defined as
\bq
D_\mu = \der_\mu - \sum_{A}\d_A (h_\mu(A))  \, .
\eq
We use ${\cal D}_\mu$ to denote a covariant derivative with respect to
$M$, $D$, U(1), SU(2) and gauge transformations\footnote{%
  However, in case of the hypermultiplet and the vector-tensor 
  multiplet we do not include the central charge transformation 
$\d_z (W_\mu^0)$ in ${\cal D}_\mu$. }.  

The composite gauge fields $\w_\mu^{ab}, \phi_\mu^i$ and
$f_\mu^a$ contained in the Weyl multiplet, are given by 
\brr \w_\mu^{ab} &=& -2e^{\nu[a}\der_{[\mu}e_{\nu]}{}^{b]}
     -e^{\nu[a}e^{b]\s}e_{\mu c}\der_\s e_\nu{}^c
     -2e_\mu{}^{[a}e^{b]\nu}b_\nu   \nonumber\\
     & & -\ft{1}{4}(2\bar{\psi}_\mu^i\g^{[a}\psi_i^{b]}
     +\bar{\psi}^{ai}\g_\mu\psi^b_i+{\rm h.c.}) \,,\nonumber\\
     \phi_\mu^i &=& (\s^{\rho\s}\g_\mu-\ft{1}{3}\g_\mu\s^{\rho\s})
     ({\cal D}_\rho\psi_\s^i-\ft{1}{8}\s\cdot T^{ij}\g_\rho\psi_{\s j}
     +\ft{1}{2}\s_{\rho\s}\chi^i) \,,\nonumber\\
     f_\mu{}^\mu &=& \ft{1}{6}{\cal R}-D
     -\Big(\ft{1}{12}e^{-1}\ve^{\mu\nu\rho\s}\bar{\psi}_\mu^i\g_\nu
     {\cal D}_\rho\psi_{\s i}
     -\ft{1}{12}\bar{\psi}_\mu^i\psi_\nu^j\,T_{ij}^{\mu\nu}
     -\ft{1}{4}\bar{\psi}_\mu^i\g^\mu\chi_i+{\rm h.c.}\Big)\,.
\label{dependent}
\err
The following supercovariant curvatures appear in the main text,
\brr \hat{R}_{\mu\nu}(Q)^i &=&
     2{\cal D}_{[\mu}\psi_{\nu]}^i
     -\g_{[\mu}\phi_{\nu]}^i
     -\ft{1}{4}\s\cdot T^{ij}\g_{[\mu}\psi_{\nu]j}\,, \nonumber\\
     \hat{R}_{\mu\nu}({\rm U(1)}) &=&
     2\der_{[\mu}A_{\nu]}
     -i\Big(\ft{1}{2}\bar{\psi}_{[\mu}^i\phi_{\nu]i}
     +\ft{3}{4}\bar{\psi}_{[\mu}^i\g_{\nu]}\chi_i-{\rm 
h.c.}\Big)\,, \nonumber\\
     \hat{R}_{\mu\nu}({\rm SU(2)})^i{}_j &=& 2\der_{[\mu} {\cal V}_{\nu ]}^i{}_j 
     + {\cal V}_{[\mu }^i{}_k{\cal V}_{\nu ]}^k{}_j \nonumber\\
     && +\Big(2\bar{\psi}_{[\mu }^i\phi_{\nu ]j} 
     -3\bar{\psi}_{[\mu }^i\g_{\nu]}\chi_j
     -({\rm h.c.}\,;\,{\rm traceless}) \Big)\,.
\err
In actual computations one may benefit from using the following 
relationships
\brr \g^\mu(\hat{R}_{\mu\nu}(Q)^i+\s_{\mu\nu}\chi^i)&=&0 \,,\nonumber\\
     2{\cal D}_{[\mu} e_{\nu]}{}^a 
     - \bar{\psi}_{[\mu}^i\g^a\psi_{\nu ]i} &=& 0\,.
\err
and 
\bq 
\begin{array}{rclrcl}
\s_{ab} &=& -\ft12\ve_{abcd}\s^{cd}\g_5 \,, \hspace{2cm} 
&\g^b\g_a\g_b &=& -2\g_a \,,\\ 
\s^{ab}\s_{ab} &=& -3 \,,&\s^{cd}\s_{ab}\s_{cd} &=& \s_{ab}\,,\\
\g^c \s_{ab}\g_c &=& 0 \,,&\s^{bc}\g_a\s_{bc} &=& 0 \,,\\
{[\g^c,\s_{ab}]} &=& 2 \d_{[a}^c \,\g_{b]} \,,
&\{\g^c,\s_{ab}\} &=& \ve_{ab}{}^{cd}\g_5\g_d \,,\\
{[\s_{ab},\s^{cd}]} &=& -4\d_{[a}{}^{[c}\s_{b]}{}^{d]}\,,
&\{\s_{ab},\s^{cd}\} &=& -\d_{[a}^c\,\d_{b]}^d
+\ft12\ve_{ab}{}^{cd}\g_5\,.
\end{array}
\eq

\end{section}



\end{document}